\documentclass[10pt]{article} % For LaTeX2e
\usepackage[preprint]{tmlr}
\usepackage{natbib}
\usepackage{graphicx}

\usepackage{amsmath,amsfonts,bm}

\def\eqref#1{equation~\ref{#1}}
\def\1{\bm{1}}

\DeclareMathAlphabet{\mathsfit}{\encodingdefault}{\sfdefault}{m}{sl}
\SetMathAlphabet{\mathsfit}{bold}{\encodingdefault}{\sfdefault}{bx}{n}

\usepackage{hyperref}
\usepackage{cleveref}
\usepackage[dvipsnames]{xcolor}
\usepackage{url}
\usepackage{enumitem}
\usepackage{booktabs}
\usepackage{tabularx}
\usepackage{array}
\usepackage{amssymb}  % for \cmark
\usepackage{pdflscape}

\newcommand{\metrictable}[5]{%
\begin{center}
\begin{tabularx}{\linewidth}{>{\centering\arraybackslash}X >{\centering\arraybackslash}X >{\centering\arraybackslash}X >{\centering\arraybackslash}X >{\centering\arraybackslash}X}
\toprule
\textbf{Extent of AIRDA} & \textbf{AI Progress} & \textbf{Oversight Capacity} & \textbf{Oversight Achieved} & \textbf{Oversight Demand} \\
\midrule
#1 & #2 & #3 & #4 & #5 \\
\bottomrule
\end{tabularx}
\end{center}
}

\hypersetup{
    colorlinks=true, 
    citecolor=NavyBlue,
    linkcolor=orange
}

\definecolor{mellowgreen}{HTML}{ADF0C7}
\definecolor{mellowred}{HTML}{FFC6C6}

\title{Measuring AI R\&D Automation}

\author{
\{Alan Chan$^{1}$\thanks{Corresponding author: \href{mailto:alan.chan@governance.ai}{alan.chan@governance.ai}} \quad
Ranay Padarath$^{1}$\thanks{Work completed as seasonal fellows at GovAI.} \quad
Joe Kwon$^{1}$\textsuperscript{\textdagger}\}\thanks{Equal contribution. Authors are free to list themselves as first author on their CVs. } \quad
Hilary Greaves$^{2}$ \quad
Markus Anderljung$^{1}$\thanks{Senior author.} \quad \\[0.5em]
{\normalfont\small 
$^1$GovAI \quad
$^2$University of Oxford}
}

\newcommand{\cmark}{\textcolor{ForestGreen}{\checkmark}}

\def\month{MM}  % Insert correct month for camera-ready version
\def\year{YYYY} % Insert correct year for camera-ready version
\def\openreview{\url{https://openreview.net/forum?id=XXXX}} % Insert correct link to OpenReview for camera-ready version

\begin{document}

\maketitle

\begin{abstract}
The automation of AI R\&D (AIRDA) could have significant implications, but its extent and ultimate effects remain uncertain. We need empirical data to resolve these uncertainties, but existing data---primarily capability benchmarks---may not reflect real-world automation or capture its broader consequences, such as whether AIRDA accelerates capabilities more than safety progress or whether our ability to oversee AI R\&D can keep pace with its acceleration. To address these gaps, this work proposes metrics to track the extent of AIRDA and its effects on AI progress and oversight. The metrics span dimensions such as capital share of AI R\&D spending, researcher time allocation, and AI subversion incidents, and could help decision makers understand the potential consequences of AIRDA, implement appropriate safety measures, and maintain awareness of the pace of AI development. We recommend that companies and third parties (e.g. non-profit research organisations) start to track these metrics, and that governments support these efforts. 
\end{abstract}

\section{Introduction}
Frontier AI companies aim to automate AI R\&D. OpenAI’s CEO Sam Altman expects to have an automated AI researcher by 2028 \citep{bellan_sam_2025}, while Google DeepMind’s CEO Demis Hassabis predicted in 2025 that an automated AI researcher was ``a few years away'' \citep{perrigo_demis_2025}. According to Anthropic's CEO Dario Amodei, current AI systems are ``good enough at coding that some of the strongest engineers [he has] ever met are now handing over almost all their coding to AI'' \citep{amodei_adolescence_2026}. Many recent products from these companies---such as Codex, AntiGravity, and Claude Code---focus on automating software engineering, a key component of AI R\&D. Adoption is widespread: 47\% of developers use AI tools daily \citep{stackoverflow_2025_2025} and 50\% of new code at Google is AI-generated \citep{alphabet_alphabet_2026}. Some qualitative reports also suggest that these tools can perform some of the R\&D tasks expected of junior AI researchers \citep{anthropic_system_2025,anthropic_system_2026}. 

\begin{figure}
    \centering
    \includegraphics[width=\linewidth]{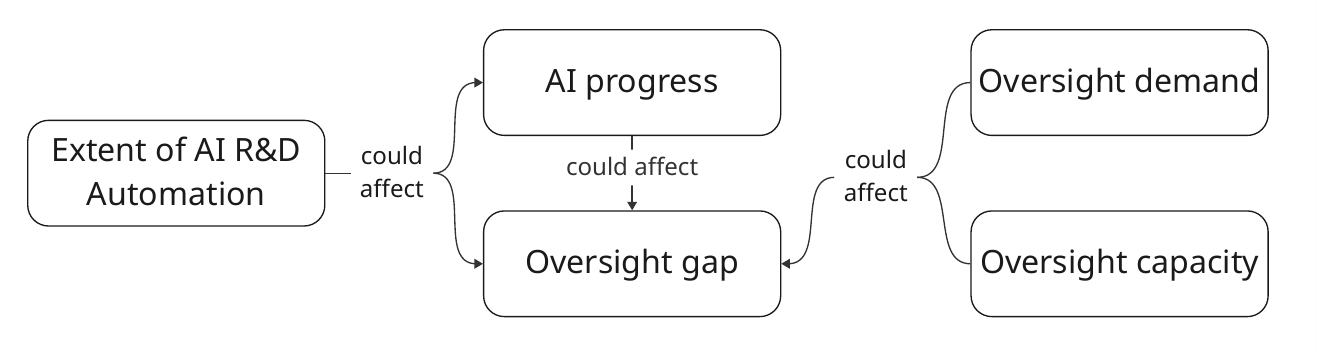}
    \caption{The \textbf{extent of AI R\&D automation} could affect both \textbf{AI progress} and the \textbf{oversight gap}: the difference between how much oversight is needed (``\textbf{oversight demand}'') and how much oversight is actually achieved. \textbf{Oversight capacity} is the ability to achieve oversight, encompassing both the ability to understand the R\&D process (e.g., having sufficient expertise) and the resources available for exercising control (e.g., human labour, monitoring tools). AI progress could also affect the oversight gap, such as by increasing the stakes of R\&D decisions. This work proposes metrics to track all of these quantities. }
    \label{fig:overall}
\end{figure}

The effects of AI R\&D automation (AIRDA) could be significant, but might cut in different directions (\Cref{fig:overall}). AIRDA could accelerate AI progress, bringing forward AI's benefits but also hastening the arrival of destructive capabilities---including those related to weapons of mass destruction---or other forms of disruption such as unemployment \citep{bengio_international_2026}. Whether this acceleration would be overall beneficial remains unclear. Key uncertainties include whether AIRDA would accelerate defensive capabilities more than offensive ones, whether safety research will keep pace with capabilities research, and whether the human institutions can adapt to the accelerated pace of progress \citep{bernardi_societal_2025,vaintrob_ai_2025,kembery_ai_2025}. 

AIRDA could also affect oversight of the AI R\&D process. By reducing the need for human researchers, AIRDA could concentrate control over AI development into fewer hands and decrease societal oversight \citep{davidson_aienabled_2025}. AI systems could also increase the need for oversight by introducing more frequent errors into the R\&D process, potentially leading to loss of control incidents \citep{benton_sabotage_2024,ward_ctrlaltdeceit_2025,bengio_international_2026}. On the other hand, AIRDA could potentially improve our ability to perform oversight, such as by enabling new AI-assisted oversight tools \citep{choi_scaling_2024}. As an analogy, higher-level programming languages let developers write more complex software that still behaves (roughly) as intended,\footnote{Compiler bugs are rare, but can still lead to subtle issues.} compared to writing programs in assembly. 

Empirical data will be essential for resolving the above uncertainties and informing decisions. Frontier AI companies may need to determine whether human review processes are keeping pace with AI-generated outputs or whether to invest more in safety research and oversight. Policymakers may need to decide what reporting requirements to impose, whether to mandate minimum levels of human involvement in AI R\&D, or how to calibrate regulatory thresholds to the actual pace of automation. 

Existing data on AIRDA has significant gaps. First, many uncertainties remain about the extent of AIRDA. Rapidly improving benchmark results indicate at least some progress \citep{jimenez_swebench_2023,chan_mlebench_2025,starace_paperbench_2025,wijk_rebench_2025,anthropic_system_2026},\footnote{At the time of writing, the most advanced model had a 80\% success rate on tasks that take human expert coders 1 hour and 10 minutes to complete \citep{metr_measuring_2025}.} but it is unclear how directly such results translate to productivity boosts given real-world integration frictions \citep{becker_measuring_2025,dellacqua_navigating_2023,brynjolfsson_generative_2023,noy_experimental_2023,narayanan_ai_2025,becker_we_2026}. Current benchmarks also mainly focus on software engineering rather than other parts of the research pipeline, such as generation and prioritization of research ideas. Second, there is little data about the consequences of AIRDA, rather than simply its extent. For example, it is unclear whether productivity improvements differ  between safety teams and other teams, or how often AI-introduced errors escape human review. 

To address these gaps, this paper proposes concrete metrics for tracking AIRDA and its potential effects on AI progress and oversight. The metrics span dimensions such as capital share of AI R\&D spending, researcher time allocation, and AI subversion incidents. Given the complexity of the underlying phenomena, these metrics are most useful in combination: no single metric perfectly correlates with AIRDA or its effects, but different metrics can compensate for each other's weaknesses and provide a broader picture. For example, a simultaneous increase in both the use of AI for high-stakes decisions (\hyperref[metric:7]{Metric \#7}) and observed defects of AI-generated R\&D outputs (\hyperref[metric:9]{Metric \#9}) could raise significant concerns. Additionally, factors such as commercial sensitivity or individual privacy could warrant sharing certain metrics only with a select group of actors, such as third-party auditors \citep{brundage_frontier_2026} or key government decision makers. 

We intend these metrics to be useful for frontier AI companies designing and implementing their safety frameworks \citep{openai_preparedness_2025,anthropic_responsible_2025,xai_xai_2025,googledeepmind_frontier_2025}, policymakers developing reporting requirements, and researchers studying AI development trajectories. 
We recommend that these actors focus on metrics (or components thereof) that are relatively neglected, which we describe in \Cref{tab:recommendations}. 

\begin{table}[h]
\caption{Recommendations for companies, governments, and third parties (e.g. non-profit research organisations).}
\label{tab:recommendations}
\centering
\begin{tabularx}{\linewidth}{l >{\raggedright\arraybackslash}p{3.2cm} X}
\toprule
\textbf{Actor} & \textbf{Recommendation} & \textbf{Metrics} \\
\midrule
Companies & Track differential progress in automating safety vs. capabilities research & \hyperref[metric:1]{Metric \#1} (AI R\&D evaluations) and \hyperref[metric:6]{Metric \#6} (staff surveys) are most straightforward. \hyperref[metric:2]{Metric \#2} (human--AI comparison RCTs) can also track differential progress, but requires significantly more effort. \\
\addlinespace
Companies & Track how AIRDA affects oversight & \hyperref[metric:7]{Metric \#7} (AI use in high-stakes decisions), \hyperref[metric:8]{Metric \#8} (researcher time allocation), and \hyperref[metric:14]{Metric \#14} (AI permission lists) are the most feasible starting points. \hyperref[metric:10]{Metric \#10} (AI subversion incidents) requires somewhat more effort, while \hyperref[metric:9]{Metric \#9} (oversight effectiveness retrospectives) requires the most effort. \\
\addlinespace
Companies & Track the actual extent of AIRDA & \hyperref[metric:11]{Metric \#11} (researcher headcount), \hyperref[metric:12]{Metric \#12} (compute distribution), and \hyperref[metric:13]{Metric \#13} (capital share) are relatively straightforward. \hyperref[metric:8]{Metric \#8} (researcher time allocation) would also be informative, but requires somewhat more implementation effort. \\
\addlinespace
Governments & Develop systems for confidential reporting, potentially in the form of industry-wide aggregates & Promising candidates are \hyperref[metric:7]{Metric \#7} (AI use in high-stakes decisions), \hyperref[metric:10]{Metric \#10} (AI subversion incidents), \hyperref[metric:12]{Metric \#12} (compute distribution), and \hyperref[metric:13]{Metric \#13} (capital share). \\
\addlinespace
Third parties & Estimate metrics using public sources & Feasible places to start are \hyperref[metric:12]{Metric \#12} (distribution of compute usage; \citealp{you_most_2025}) and \hyperref[metric:11]{Metric \#11} (researcher headcount). \\
\addlinespace
Third parties & Create tooling and design surveys & Time-tracking software for \hyperref[metric:8]{Metric \#8} (researcher time allocation), and surveys for \hyperref[metric:6]{Metric \#6} (self-reported productivity boosts) and \hyperref[metric:7]{Metric \#7} (AI use in high-stakes decisions). \\
\bottomrule
\end{tabularx}
\end{table}

\section{What is AI R\&D?}
AI R\&D encompasses the activities involved in developing and improving AI systems, including both capabilities research and safety and security research. Roughly following \citet{owen_interviewing_2024}, we explain activities involved in the current AI R\&D process and comment on how AI could plausibly automate its various stages. 

\textbf{Coming up with and prioritizing research ideas}. Research ideas can cover a wide range of topics, such as how an AI system might behave on a new evaluation, whether a particular technique might improve performance, or what the cause of a behaviour might be. AI systems could potentially help generate, filter, and prioritize such ideas. As an early experiment, \citet{wen_predicting_2025} show that AI systems can predict which of two research ideas will perform better on benchmarks, in some cases outperforming human experts. 

\textbf{Designing experiments}. Ideas need to be tested in experiments. Designing these experiments involves tasks such as writing and debugging code, as well as generating or collecting datasets (e.g., for training or evaluation). Coding agents have made significant gains in the past few years, with improved performance on coding benchmarks \citep{jimenez_swebench_2023,chan_mlebench_2025} and widespread adoption of tools like GitHub Copilot, Cursor, and Claude Code among developers \citep{stackoverflow_2025_2025}. Synthetically generated data is also becoming increasingly important to the AI R\&D pipeline \citep{bai_constitutional_2022,schoen_stress_2025,fronsdal_petri_2025}. 

\textbf{Running experiments}. Running experiments can sometimes be as simple as running the code. Other times, especially for large-scale experiments, AI researchers have to efficiently distribute experiments across multiple machines and monitor experiments for signs of failure. Monitoring large-scale training runs can be particularly tricky, involving heuristics and on-the-fly interventions \citep{marin_marin_2025}. AI systems can potentially assist with monitoring and troubleshooting training runs, but data on effectiveness remains limited. 

\textbf{Analyzing results}. Analysis of experimental results can inform new research ideas and experiments. AI systems already assist with aspects of result analysis, such as generating visualizations and identifying patterns across runs \citep{choi_scaling_2024,fronsdal_petri_2025}. 

We use \textbf{AI R\&D automation (AIRDA)} to refer to the use of AI to carry out parts of this pipeline. Automation can be implemented to differing degrees, from simply using AI as a hypothesis generator, to deploying teams of artificial researchers that carry out all parts of the pipeline. The tasks involved in AI R\&D could also change over time. For example, increasing involvement of AI in the R\&D pipeline could shift human involvement towards verifying that experiments have been designed and run correctly. 

\section{Potential Implications of AI R\&D Automation}\label{sec:implications}
\begin{figure}[!htb]
    \centering
    \includegraphics[width=\linewidth]{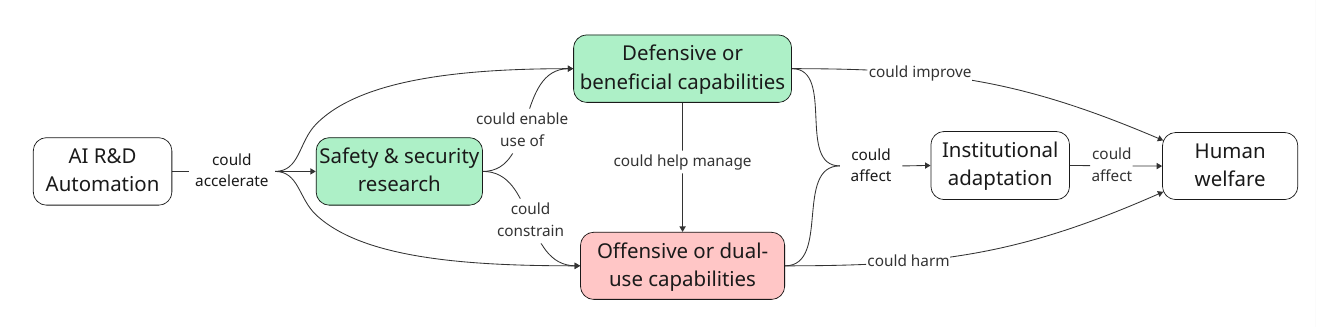}
    \caption{\colorbox{mellowgreen}{Positive} and \colorbox{mellowred}{negative} implications of AI R\&D automation for AI progress. }
    \label{fig:ai-progress}
\end{figure}

\begin{figure}
    \centering
    \includegraphics[width=0.8\linewidth]{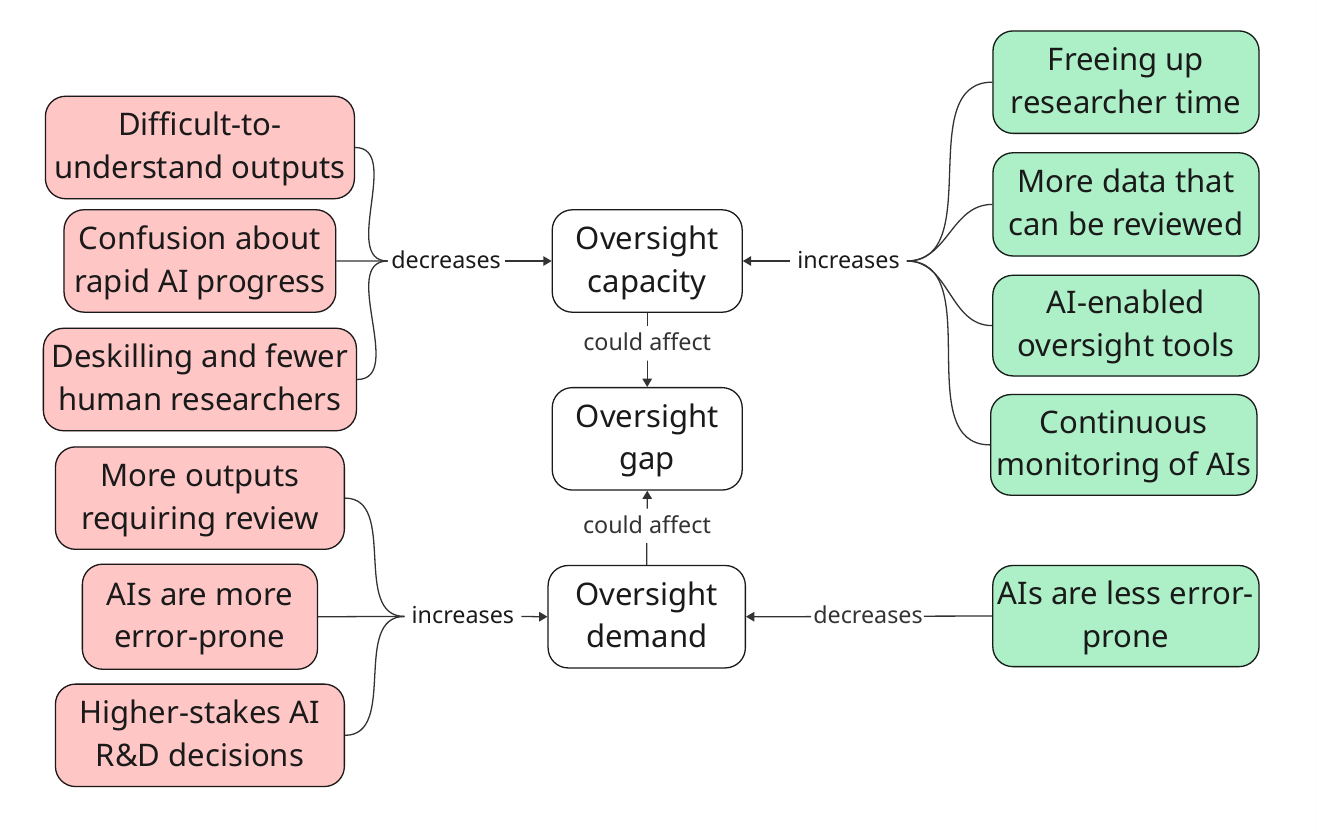}
    \caption{%The oversight gap (not shown) is the difference between how much oversight is needed for a given level of safety and/or assurance (``\textbf{oversight demand}'') and how much oversight is actually performed (``\textbf{level of oversight}''). \textbf{Oversight capacity} is the ability to perform oversight, and affects the level of oversight. 
    AI R\&D automation could \colorbox{mellowgreen}{decrease} or \colorbox{mellowred}{increase} the oversight gap by changing oversight capacity, the level of oversight achieved (not depicted for simplicity), or oversight demand. Note that oversight capacity may not directly translate into oversight achieved (e.g., companies decide to forgo oversight due to financial or time costs). 
    }
    \label{fig:oversight}
\end{figure}

To identify informative metrics for measuring AIRDA, we first discuss its potential implications for AI progress and oversight. For AI progress, AIRDA could affect when certain kinds of capabilities emerge and how we respond to them (\Cref{fig:ai-progress}). For oversight, AIRDA could affect the gap between how much oversight of AI R\&D is needed and how much is achieved (\Cref{fig:oversight}).

\subsection{AI Progress}

AIRDA could accelerate both capabilities research and safety \& security research. One potential mechanism is a recursive feedback loop, where AI systems generate software improvements that lead to more capable AI systems, which in turn generate further improvements \citep{eth_will_2025}. Numerous copies of automated AI researchers could run in parallel as long as there is available compute, working around the clock and much faster than humans can. In the long run, AI systems could become more capable than human researchers and generate higher-quality improvements \citep{ho_algorithmic_2024,barnett_algorithmic_2025,sevilla_compute_2022}. While empirical evidence for these scenarios is limited, early signs are suggestive: AI systems already write code much faster than humans can, and can in limited cases outperform human experts in predicting which idea will lead to better performance on benchmarks \citep{wen_predicting_2025}. Compute bottlenecks \citep{erdil_most_2025,whitfill_will_2025,ho_least_2026} and integration frictions could dampen acceleration, but at least some speed-up seems plausible. 

Such acceleration could have significant consequences, including bringing forward AI’s benefits, but whether it would be overall beneficial remains unclear. The reason is that AIRDA-induced acceleration %would not simply accelerate the whole world, but rather 
could (i) accelerate some aspects of AI progress relative to others, and (ii) accelerate AI progress without similarly accelerating crucial human and institutional processes. The order in which technologies arrive can have significant consequences \citep{sandbrink_differential_2022,buterin_my_2023,nielsen_notes_2024}.

First, AIRDA could affect whether offensive AI capabilities arrive before the defensive capabilities or safety and security mitigations needed to counter them. For example:
\begin{itemize}
    \item AI forecasting capabilities could help to manage CBRN uplift capabilities, by predicting when and how malicious actors might carry out attacks. If forecasting is easier to automate than CBRN capabilities, then AIRDA would yield a net benefit by providing stronger defences at any given level of CBRN uplift. %If the reverse holds, AIRDA would yield a net harm in this case. 
    \item %Some safety tasks---such as red-teaming---may be particularly well-suited to automation \citep{fronsdal_petri_2025,fronsdal_petri_2026,gupta_bloom_2025}, in which case AIRDA may yield a net benefit. On the other hand, 
    Safety and security research could facilitate the use of defensive capabilities (e.g., by ensuring that systems follow user instructions) or directly constrain offensive capabilities (e.g., through misuse safeguards). However, safety may be harder to accelerate than capabilities because safety is more difficult to measure. For example, attempts from AI systems to hide misbehaviour could make it hard to assess whether safety is actually improving \citep{greenblatt_alignment_2024,meinke_frontier_2025}. If so, accelerated capabilities progress could outstrip the development of necessary safeguards, which could result in a net harm.  
\end{itemize}
Even if AIRDA differentially accelerates offensive capabilities, the resulting equilibrium could be stable if using the capability results in catastrophic harm. For example, some scholars argue that the spread of nuclear weapons has improved peace through deterrence \citep{waltz_spread_1981}, though others dispute this claim \citep{sagan_spread_2003}. An additional nuance is that many capabilities are dual-use: for example, AI systems that identify software vulnerabilities could strengthen both defensive cybersecurity and offensive attacks. In such cases, a key question is whether AIRDA affects the extent to which defensive applications can be deployed before harmful uses proliferate. %If AIRDA is initially concentrated within AI companies, for instance, they could help other actors strengthen their cyber defenses before widely releasing models.

Second, human and institutional responses to advanced AI operate on timescales that may not accelerate alongside AI progress itself. Understanding a new technology's effects and developing appropriate responses already takes years; accelerated capability gains could leave institutions even less prepared than they are now. For instance, accelerated AI progress could displace workers faster than they can retrain and adapt \citep{korinek_scenarios_2024,manning_how_2026,bengio_international_2026}. Rapid progress could also push decision-makers toward hasty or poorly-informed responses, such as overly broad or unenforceable restrictions on AI development. That said, accelerated progress could also galvanize policy attention and reforms, and advances in defensive capabilities like forecasting could aid adaptation \citep{schoenegger_aiaugmented_2024,karger_forecastbench_2025}.

\subsection{Oversight}\label{sec:oversight}
% \textbf{Oversight} is the act of informed supervision and control of the AI R\&D process to produce intended results, through activities such as reviewing AI-generated outputs for errors and maintaining control over research direction. 

An actor has \textbf{oversight} over the AI R\&D process to the extent that they (1) understand the process and (2) exercise informed control over it in order to produce desired outputs, such as by reviewing AI-generated outputs for errors. In other words, oversight is about both a particular state (i.e., understanding) and a particular act (i.e., exercising control). An actor performs oversight to the extent that they exercise such control. 

We ultimately care about the \textbf{oversight gap}: the difference between how much oversight is needed over the AI R\&D process (``\textbf{oversight demand}'')\footnote{Oversight could improve the safety of and provide assurance about the AI R\&D process, but could, for example, impact research productivity or impose financial or other costs. Different actors could have different preferences about how to make these trade-offs. To sidestep this issue, we focus on changes to oversight demand rather than its absolute level.} and how much oversight is actually achieved. A factor that could affect the gap is \textbf{oversight capacity}: the ability to achieve oversight, encompassing both 
% the resources available for oversight (e.g., human labour) and the cost-effectiveness of oversight methods.\footnote{We combine these two distinct concepts for simplicity.} 
the ability to understand the R\&D process (e.g., having sufficient expertise) and the resources available for exercising control (e.g., human labour, monitoring tools). 
Crucially, oversight capacity does not directly translate into oversight achieved: factors such as financial costs or negligence could lead organizations to forgo oversight. 

We analyse how AIRDA could affect the oversight gap by changing oversight capacity, the level of oversight achieved, or oversight demand. 
% The net effect of AIRDA on the balance between oversight capacity and demand remains unclear. 

\subsubsection{Decreasing the Oversight Gap}
AIRDA could decrease the oversight gap by increasing oversight capacity or the level of oversight achieved, or decreasing oversight demand. 

\textbf{Increasing oversight capacity or the level of oversight achieved}:
\begin{itemize}
    \item By automating routine tasks such as writing code or running experiments, AIRDA could allow human researchers to spend more time on oversight.\footnote{In practice, the time gained could be spent advancing capabilities, which could increase oversight demand by accelerating AI progress.} 
    \item AI systems generate extensive logs, chains of thought, activations, and other intermediate outputs that make their work more transparent and auditable than equivalent human work.
    \item AIRDA could accelerate the development of automated AI oversight tools \citep{choi_scaling_2024,fronsdal_petri_2025,fronsdal_petri_2026}, which could be more effective than humans at carrying out oversight or could do so more cheaply. 
    \item Continuous monitoring is not practical or socially acceptable for humans, but is feasible for AI agents and is actively being explored \citep{openai_gpt53codex_2026,anthropic_responsible_2026}.
\end{itemize}

\textbf{Decreasing oversight demand}: 
\begin{itemize}
    \item More capable AI systems could be less error-prone than humans.
\end{itemize}

\subsubsection{Increasing the Oversight Gap}
AIRDA could increase the oversight gap by decreasing oversight capacity or the level of oversight achieved, or increasing oversight demand. 

\textbf{Decreasing oversight capacity or the level of oversight achieved}: 
\begin{itemize}
    \item The results of automated AI research could potentially be more difficult for humans to understand (e.g., results might use novel mathematics), or AI-introduced errors could be more difficult for humans to notice. %Because more effort might be needed to achieve the same level of oversight, oversight capacity would decrease. 
    \item AIRDA could motivate companies to reduce the number of human researchers involved in AI R\&D. As a result, there might be less human labour available to carry out oversight activities.
    \item AIRDA could reduce hands-on involvement in AI R\&D. If so, human researchers could lose the detailed, ground-level understanding of systems and codebases needed to identify errors. There might also be reduced oversight in practice (e.g., giving instructions to coding agents without supervising their activities), which would decrease the level of oversight achieved.
    \item Accelerated AI progress could make it difficult for humans to understand the AI R\&D process and make informed decisions (e.g., time pressure leads to carelessness).
    \item If AIRDA results in radically fewer human researchers, decisions to develop and deploy potentially dangerous capabilities could concentrate into the hands of a handful of humans with limited internal or external checks \citep{macaskill_preparing_2025}, leading to reduced oversight at a societal level.
\end{itemize}

\textbf{Increasing oversight demand}:
\begin{itemize}
    \item AI systems could be more error-prone than humans \citep{cotroneo_humanwritten_2025}, for reasons including capability limitations, adversarial interference (e.g., poisoning training data), or reward hacking \citep{betley_emergent_2026,macdiarmid_natural_2025}. 
    \item Automated researchers can work faster than humans and generate a larger volume of experiments, code, and decisions requiring review per unit time.
    \item If AIRDA results in faster AI progress, individual R\&D decisions could be higher stakes. For example, a decision to develop the next model could result in much larger capability jumps than occur today.
    % \item If AIRDA results in radically fewer human researchers, decisions to develop and deploy potentially dangerous capabilities could concentrate into the hands of a handful of humans with limited internal or external checks \citep{macaskill_preparing_2025}.
\end{itemize}

\section{Metrics for AI R\&D Automation }\label{sec:metrics}

We propose metrics to track the extent of AI R\&D automation, the rate of AI progress, and changes in oversight, focusing on the uncertainties identified in the previous section.

For the \textbf{extent of AIRDA}, the metrics track:
\begin{itemize}
    \item Capability evaluations as leading indicators of whether AI R\&D could be automated (Metrics \hyperref[metric:1]{\#1}, \hyperref[metric:2]{\#2})
    \item The extent of AI R\&D automation in practice (Metrics \hyperref[metric:6]{\#6}, \hyperref[metric:7]{\#7}, \hyperref[metric:8]{\#8})
    \item Other organizational and operational indicators of AI R\&D automation (Metrics \hyperref[metric:11]{\#11}, \hyperref[metric:12]{\#12}, \hyperref[metric:13]{\#13}, \hyperref[metric:14]{\#14})
\end{itemize}

For \textbf{AI progress}, the metrics track both the overall rate of AI progress and the rate of safety progress relative to capabilities progress (Metrics \hyperref[metric:1]{\#1}, \hyperref[metric:2]{\#2}, \hyperref[metric:5]{\#5}, \hyperref[metric:6]{\#6}, \hyperref[metric:11]{\#11}). 

For \textbf{oversight}, we track oversight capacity, how much oversight is actually achieved, and oversight demand (the latter two of which give us the oversight gap). %$, and . For oversight gap, some metrics separately track oversight demand or how much oversight is actually performed, while others track oversight gap directly. 
More specifically, the metrics track:
\begin{itemize}
    \item How human researchers spend their time (Metrics \hyperref[metric:6]{\#6}, \hyperref[metric:8]{\#8})
    \item The extent to which AI systems help with oversight (Metrics \hyperref[metric:3]{\#3})
    \item Whether AI systems act in ways that increase the demand for oversight (Metrics \hyperref[metric:4]{\#4}, \hyperref[metric:9]{Metric \#9}, \hyperref[metric:10]{\#10})
    \item Headcount and performance of human researchers (\hyperref[metric:11]{Metric \#11})
    \item The extent to which AI systems are involved in significant decisions or actions (Metrics \hyperref[metric:7]{\#7}, \hyperref[metric:14]{\#14})
\end{itemize}

Many of these metrics are most useful in combination. For example, an increase in the use of AI for high-stakes decisions (\hyperref[metric:7]{Metric \#7}), coupled with an increase in the defects of AI-generated R\&D outputs (\hyperref[metric:9]{Metric \#9}), could raise significant concerns.

In the rest of this section, we describe what each metric tracks and analyze feasibility, limitations, and information sensitivity considerations. Sensitivity considerations could warrant sharing certain metrics only with a limited group of actors, such as key government decision makers. We summarize the metrics in \Cref{tab:experimental-metrics,tab:survey-metrics,tab:operational-metrics,tab:organizational-metrics}. 

\begin{landscape}
\begin{table}
\centering
\caption{Experimental metrics for measuring AI R\&D automation.}
\label{tab:experimental-metrics}
\footnotesize
\renewcommand{\arraystretch}{1.4}
\begin{tabularx}{0.95\linewidth}{>{\raggedright\arraybackslash}p{3cm} >{\raggedright\arraybackslash}p{4cm} >{\raggedright\arraybackslash}X >{\raggedright\arraybackslash}X >{\raggedright\arraybackslash}X >{\raggedright\arraybackslash}X >{\raggedright\arraybackslash}X >{\raggedright\arraybackslash}X}
\toprule
\textbf{Metric} & \textbf{Description} & \textbf{Extent of AIRDA} & \textbf{AI Progress} & \textbf{Oversight Capacity} & \textbf{Oversight Achieved} & \textbf{Oversight Demand} & \textbf{What remains to be done} \\
\midrule
\hyperref[metric:1]{Metric \#1}: AI performance on AI R\&D evaluations & AI performance on evaluations of capabilities directly relevant to AI R\&D work, such as replicating ML papers and picking research ideas to test. & \cmark~Leading indicator of AI systems' potential to perform AI R\&D. & \cmark~Can compare improvements on safety research tasks to capabilities research tasks. & & & & Developing more evaluations for non-software-engineering tasks and safety and security research tasks. \\
\midrule
\hyperref[metric:2]{Metric \#2}: AI performance on AI R\&D evaluations compared to humans and human-AI teams (``AI R\&D Performance RCTs'') & Performance differences between human-AI teams, human-only teams, and AI-only teams on AI R\&D tasks. & \cmark~Superior AI-only performance suggests greater AIRDA potential. & \cmark~Can compare improvements on safety research tasks to capabilities research tasks. & & & \cmark~Depending on the interaction protocol, superior human-AI performance can suggest non-negligible oversight demand. & Broadening coverage of AI R\&D tasks and running more comparisons between AI systems and human-AI teams.\\
\midrule
\hyperref[metric:3]{Metric \#3}: Oversight red-teaming experiments & Evaluations for whether oversight systems used in production would catch an AI system that is directed to subvert the R\&D process, such as by attempting to sabotage experiments. & & & \cmark~Measures ability to catch attempts to subvert the AI R\&D process. & & & Developing and running evaluations specific to internal infrastructure. \\
\midrule
\hyperref[metric:4]{Metric \#4}: Misalignment evaluations & Evaluations measuring propensities and capabilities relevant to misaligned behaviour, including alignment faking, sycophancy, sabotage, and reward hacking. & & & & & \cmark~Greater misalignment increases oversight demand. & Developing evaluations that are more representative of realistic AI R\&D workflows. \\
\midrule
\hyperref[metric:5]{Metric \#5}: Compute efficiency improvements & The year-over-year percentage reduction in the amount of compute needed to achieve a given level of performance on capability (including AI R\&D) evaluations. & \cmark~Can measure efficiency improvements on AI R\&D tasks. & \cmark~Directly tracks AI progress. & & & & Calculating efficiency improvements from existing data. Potentially running additional evaluations. \\
\bottomrule
\end{tabularx}
\end{table}
\end{landscape}

\begin{landscape}
\begin{table}
\centering
\caption{Survey-based metrics for measuring AI R\&D automation.}
\label{tab:survey-metrics}
\footnotesize
\renewcommand{\arraystretch}{1.4}
\begin{tabularx}{0.95\linewidth}{>{\raggedright\arraybackslash}p{3cm} >{\raggedright\arraybackslash}p{4cm} >{\raggedright\arraybackslash}X >{\raggedright\arraybackslash}X >{\raggedright\arraybackslash}X >{\raggedright\arraybackslash}X >{\raggedright\arraybackslash}X >{\raggedright\arraybackslash}X}
\toprule
\textbf{Metric} & \textbf{Description} & \textbf{Extent of AIRDA} & \textbf{AI Progress} & \textbf{Oversight Capacity} & \textbf{Oversight Achieved} & \textbf{Oversight Demand} & \textbf{What remains to be done} \\
\midrule
\hyperref[metric:6]{Metric \#6}: Staff views on AI use and productivity boosts & Researchers' self-reported AI use patterns and productivity gains across different AI R\&D task categories. & \cmark~Elicits self-reported AI usefulness across R\&D tasks. & \cmark~Can compare productivity boosts across teams (e.g.\ safety vs.\ pre-training teams). & \cmark~Can ask how helpful AI is for oversight tasks. & & & Designing questions and establishing a regular cadence for running surveys. \\
\midrule
\hyperref[metric:7]{Metric \#7}: Extent of AI use in high-stakes decisions & Survey of how AI is involved in high-stakes R\&D decisions, such as research agenda prioritization, training runs, and deployment decisions. & \cmark~Self-reported signal of AI involvement in key decisions. & & & \cmark~Can ask about how much oversight is exercised over AI involvement in high-stakes decisions. & \cmark~Significant AI involvement suggests greater oversight demand. & Identifying high-stakes R\&D decisions. Designing surveys and running them each time such decisions are made. \\
\bottomrule
\end{tabularx}
\end{table}
\end{landscape}

\begin{landscape}
\begin{table}
\centering
\caption{Operational metrics for measuring AI R\&D automation.}
\label{tab:operational-metrics}
\footnotesize
\renewcommand{\arraystretch}{1.4}
\begin{tabularx}{0.95\linewidth}{>{\raggedright\arraybackslash}p{3cm} >{\raggedright\arraybackslash}p{4cm} >{\raggedright\arraybackslash}X >{\raggedright\arraybackslash}X >{\raggedright\arraybackslash}X >{\raggedright\arraybackslash}X >{\raggedright\arraybackslash}X >{\raggedright\arraybackslash}X}
\toprule
\textbf{Metric} & \textbf{Description} & \textbf{Extent of AIRDA} & \textbf{AI Progress} & \textbf{Oversight Capacity} & \textbf{Oversight Achieved} & \textbf{Oversight Demand} & \textbf{What remains to be done} \\
\midrule
\hyperref[metric:8]{Metric \#8}: Researcher time allocation (``AI-powered Toggl'') & How human researchers allocate time across different R\&D activities, measured via automated time tracking software. & \cmark~Captures potential impacts of AIRDA on researcher time allocation. & & & \cmark~Can measure time spent on oversight activities. & & Developing and deploying time-tracking software, managing privacy concerns. \\
\midrule
\hyperref[metric:9]{Metric \#9}: Oversight effectiveness retrospectives & The proportion of AI-generated R\&D outputs (e.g.\ production code, experimental analyses) that have defects, stratified by level of oversight. & & &  & \cmark~Tracks oversight activities on outputs. & \cmark~Tracks oversight gap directly, and oversight demand indirectly if level of oversight is held constant. & Building a tracking/tagging pipeline and establishing a regular data analysis schedule. \\
\midrule
\hyperref[metric:10]{Metric \#10}: AI subversion incidents & The number and severity of incidents in which AI systems attempted to subvert real AI R\&D processes, such as by sabotaging experiments or hiding misbehaviour. & & & & & \cmark~An increasing rate of subversion incidents indicates growing oversight demand. & Defining what counts as a subversion incident and building detection infrastructure. \\
\bottomrule
\end{tabularx}
\end{table}
\end{landscape}

\begin{landscape}
\begin{table}
\centering
\caption{Organizational metrics for measuring AI R\&D automation.}
\label{tab:organizational-metrics}
\footnotesize
\renewcommand{\arraystretch}{1.4}
\begin{tabularx}{0.95\linewidth}{>{\raggedright\arraybackslash}p{3cm} >{\raggedright\arraybackslash}p{4cm} >{\raggedright\arraybackslash}X >{\raggedright\arraybackslash}X >{\raggedright\arraybackslash}X >{\raggedright\arraybackslash}X >{\raggedright\arraybackslash}X >{\raggedright\arraybackslash}X}
\toprule
\textbf{Metric} & \textbf{Description} & \textbf{Extent of AIRDA} & \textbf{AI Progress} & \textbf{Oversight Capacity} & \textbf{Oversight Achieved} & \textbf{Oversight Demand} & \textbf{What remains to be done} \\
\midrule
\hyperref[metric:11]{Metric \#11}: Headcount of AI researchers and the distribution of their performance & The number and seniority of AI researchers at frontier AI companies, along with the distribution of their performance, as estimated, for example, by salaries or compute budgets. & \cmark~Decreased headcount suggests increased AIRDA. & \cmark~Useful AI tools could lead to a widening performance distribution. & \cmark~Decreased headcount or performance could reduce oversight capacity. & \cmark~Decreased headcount could reduce how much oversight is performed. & & Aggregating and presenting existing data. \\
\midrule
\hyperref[metric:12]{Metric \#12}: Distribution of compute usage & How compute is distributed across various AI R\&D activities. & \cmark~AIRDA could shift compute use towards internal inference. & & & & & Developing systems to classify compute use. \\
\midrule
\hyperref[metric:13]{Metric \#13}: Capital share of AI R\&D spending & Compute expenditure as a percentage of total AI R\&D spending. & \cmark~Rising capital-to-labour ratio could signal increasing automation. & & & & & Calculating and reporting the metric. \\
\midrule
\hyperref[metric:14]{Metric \#14}: AI permission lists & A list of which actions AI systems are authorized to take with different levels of human approval or review, including where none is required. & \cmark~Automation is potentially higher if AI systems are permitted to take significant actions autonomously. & & &  & \cmark~Oversight demand may be higher if AI systems are permitted to take significant actions without human review. & Defining actions of interest, collecting permissions across teams, and updating the list regularly. \\
\bottomrule
\end{tabularx}
\end{table}
\end{landscape}

\subsection{Experimental Metrics}\label{sec:experimental}
These metrics involve running experiments on AI systems. 

\subsubsection{Metric \#1: AI performance on AI R\&D evaluations}\label{metric:1}

\metrictable{\cmark}{\cmark}{}{}{}

\textbf{Description}: This metric tracks capabilities directly relevant to AI R\&D work, including software engineering, ML experimentation, research replication, and idea generation and curation. Current evaluations mostly cover tasks related to software engineering, and include SWE-Bench, MLE-Bench, RE-BENCH, and PaperBench \citep{jimenez_swebench_2023,chan_mlebench_2025,wijk_rebench_2025,starace_paperbench_2025}. Where available, both AI-only performance and human baseline comparisons should be reported. Additionally, if the time humans took to complete tasks is available, the $x$\%-task-completion time horizon for various $x$ (e.g., 50\%, 80\%) should be reported (see~\citet{metr_measuring_2025} for a more precise definition).

\textbf{Significance}: Capability evaluations provide information about the extent to which AI systems could automate core AI R\&D activities. Because certain tasks could be more relevant to safety research than to capabilities research, tracking performance across different task types could offer insight into differential progress across research areas.

\textbf{Feasibility}: High, especially since benchmark results are commonly reported \citep{anthropic_system_2025,openai_gpt5_2025,googledeepmind_gemini_2025}. The main costs are evaluation design, compute for running the evaluation, and engineering time (e.g., testing different agent scaffolds). 

\textbf{Limitations}:
\begin{itemize}
    \item \textbf{Ecological validity}: Benchmarks do not fully capture R\&D work, which involves ambiguous objectives, longer time horizons, coordination across projects, and messier feedback loops. As such, raw performance numbers can tell us that we are getting closer to AIRDA, but might not tell us how far away we are. 
    \item \textbf{Data contamination}: Some evaluations are drawn from public sources (e.g., GitHub, Kaggle, published papers) and could therefore leak into training sets. As a result, this metric could overestimate AI R\&D capabilities. 
    \item \textbf{Autonomous evaluations only}: These benchmarks evaluate AI agents working independently, but human-AI collaborative teams may better reflect how AI will actually be integrated into R\&D work (see \hyperref[metric:2]{Metric \#2}).
    \item \textbf{Setup sensitivity}: Results can depend on factors such as the task setup and agent scaffolding, making comparisons across setups difficult~\citep{zhu_establishing_2025}.
    \item \textbf{Saturation}: Evaluations can become quickly saturated. For example, Claude Opus 4.6 has saturated all of Anthropic's cyber evaluations \citep{anthropic_system_2026}.
\end{itemize}

\textbf{Sensitivity}: Low. Organizations with strong performance will likely share results voluntarily for marketing and recruiting, but weak results could be omitted or selectively reported.

\textbf{What remains to be done}: Developing more evaluations for tasks not related to software engineering, such as idea generation and curation, and for tasks that may be more relevant to safety and security research than capabilities research (e.g., interpretability, red-teaming).

\subsubsection{Metric \#2: AI performance on AI R\&D evaluations compared to humans and human-AI teams (``AI R\&D Performance RCTs'')}\label{metric:2}

\metrictable{\cmark}{\cmark}{}{}{\cmark}

\textbf{Description}: This metric measures performance differences between human-AI teams, human-only teams, and AI-only teams on AI R\&D tasks. The human-AI collaborative teams should have predefined interaction protocols (e.g., humans can review AI outputs, redirect approaches, provide context, or take over subtasks). Both absolute performance in each group and the gap between groups should be reported.

\textbf{Significance}: Superior performance of AI-only teams would indicate greater potential for AIRDA. Superior human-AI performance would indicate that humans still provide value to AI R\&D, but more precise implications would depend upon the human-AI interaction protocol. For example, if the protocol mostly involves a human overseeing AI work, superior human-AI performance would indicate that human oversight still meaningfully improves outputs. On the other hand, we would not obtain information about oversight if humans and AI systems mostly complete tasks in parallel without interacting.

\textbf{Feasibility}: Moderate. Although some organizations have begun conducting related studies~\citep{becker_measuring_2025,becker_we_2026,dellacqua_navigating_2023,brynjolfsson_generative_2023,noy_experimental_2023,metr_measuring_2025}, human-AI evaluation requires substantial effort and cost, including recruiting human participants with relevant expertise (expensive for skilled ML or SWE tasks), designing standardized collaboration protocols, and running parallel AI-only and human-AI evaluations on matching task sets. However, where a human expert baseline already exists, the additional cost of comparing it to an AI-only evaluation is minimal. Furthermore, a lighter-weight, but less objective, version of this metric might instead be based on researchers' self-reports rather than experiments (see \hyperref[metric:6]{Metric \#6}).

\textbf{Limitations}:
\begin{itemize}
    \item \textbf{Problems with evaluations}: Because this metric is also based on evaluations, it inherits many of the same limitations as \hyperref[metric:1]{Metric \#1}, such as ecological validity and setup sensitivity. 
    % \item \textbf{Protocol sensitivity}: Results depend heavily on how human-AI collaboration is structured. Different interaction patterns (e.g.\ human-in-the-loop vs.\ human-on-the-loop, review frequency, veto authority) may yield different gaps.
    \item \textbf{Comparability over time}: Human participants have different skills and human performance is less reproducible than AI performance. As such, it could be difficult to compare RCT results over time. 
    \item \textbf{Dependence on human expertise levels}: The gap between teams will vary with human expertise level. Results should clearly specify the skill level of human participants.
    \item \textbf{Performance gaps likely depend on task selection}: Some tasks inherently favor collaboration (ambiguous, long-horizon problems), while others favor autonomous execution (well-defined, parallelizable work). The metric is most informative when tracked across diverse task types.
\end{itemize}

\textbf{Sensitivity}: Low to moderate, depending on the results. Results showing that human-AI teams substantially outperform AI-only systems could undermine marketing claims about AI agents. Moreover, superior performance of AI-only systems could motivate public concerns about unemployment or a potentially uncontrollable intelligence explosion~\citep{macaskill_preparing_2025}.

\textbf{What remains to be done}: Covering a broader array of AI R\&D tasks in future studies. In principle, every benchmark used in \hyperref[metric:1]{Metric \#1} could also be used for this metric.

\subsubsection{Metric \#3: Oversight red-teaming experiments}\label{metric:3}

\metrictable{}{}{\cmark}{}{}

\textbf{Description}: In controlled experiments, this metric measures how often an organization's oversight systems would catch an AI system explicitly instructed to subvert normal R\&D workflows, such as by attempting to sabotage experiments, insert backdoors, or manipulate evaluation results. The metric is reported as the percentage of subversive actions that are detected, both overall and by task category.

\textbf{Significance}: All else equal, oversight systems that are more capable of detecting subversion attempts in test deployments will likely be more effective during the real R\&D process. However, because this metric involves controlled experiments, it does not quite measure the level of oversight achieved. 

\textbf{Feasibility}: Low to moderate. Implementation requires a sandboxed R\&D environment mirroring production workflows, models fine-tuned or prompted to attempt specific behaviours, and instrumentation to log both attempted actions and detection events. Some related evaluations exist \citep{inglis_controlarena_2025}, but may not necessarily reflect real production workflows. A related line of work uses red teams to embed hidden objectives in a language model, with blue teams attempting to detect them \citep{marks_auditing_2025}.

\textbf{Limitations}:
\begin{itemize}
    \item \textbf{Does not measure propensity}: The metric does not capture whether AI systems would attempt subversion unprompted. Prompting AI systems to be subversive might produce behaviours that would not surface during normal use of AI for R\&D.
    \item \textbf{Reduced realism}: Testing must occur in sandboxed environments to prevent actual harm, which may reduce realism and potentially affect model behaviour.
\end{itemize}

\textbf{Sensitivity}: Low to moderate. Aggregate results would be less sensitive than breakdowns that reveal specific attack vectors. Organizations with strong detection rates might selectively share results to demonstrate security/oversight maturity. 

\textbf{What remains to be done}: Making evaluations specific to internal company infrastructure.

\subsubsection{Metric \#4: Misalignment evaluations}\label{metric:4}

\metrictable{}{}{}{}{\cmark}

\textbf{Description}: This metric measures AI systems' propensities and capabilities relevant to misaligned behaviour in AI R\&D environments. Misaligned behaviours could include alignment faking~\citep{greenblatt_alignment_2024}, scheming~\citep{meinke_frontier_2025}, sabotage~\citep{benton_sabotage_2024,ward_ctrlaltdeceit_2025}, and reward hacking~\citep{betley_emergent_2026}.

\textbf{Significance}: Greater misalignment increases oversight demand.

\textbf{Feasibility}: High. Many AI companies and external organizations already run such evaluations \citep{openai_gpt5_2025,anthropic_system_2025,googledeepmind_gemini_2025,ward_ctrlaltdeceit_2025,betley_emergent_2026,macdiarmid_natural_2025}.

\textbf{Limitations}:
\begin{itemize}
    \item \textbf{Lack of generalizability}: Differences between evaluation environments and real AI R\&D workflows could limit the informativeness of evaluations. For example, real R\&D workflows likely involve lengthier, more complex interaction patterns than can be simulated in an evaluation. 
    \item \textbf{Evaluation gaming}: Data leakage or attempts to train on evaluations could make this metric less informative.
    \item \textbf{Evaluation awareness}: Frontier models increasingly recognize when they are being evaluated and condition their behaviour accordingly~\citep{meinke_frontier_2025}. As a result, misalignment evaluations could underestimate the degree of misalignment in practice.
\end{itemize}

\textbf{Sensitivity}: Low to moderate. As noted above, several AI companies already publish misalignment evaluation results in system cards. However, high misalignment scores in the future could reflect poorly on their products.

\textbf{What remains to be done}: Developing evaluations that are more representative of realistic AI R\&D workflows.

\subsubsection{Metric \#5: Compute efficiency improvements}\label{metric:5}

\metrictable{\cmark}{\cmark}{}{}{}

\textbf{Description}: This metric measures the year-over-year percentage reduction in the amount of compute (FLOP) needed to achieve a given level of performance on capabilities (including AI R\&D) evaluations. For example, if achieving a fixed benchmark score requires half as much compute this year as last year, that represents a 2x improvement in compute efficiency. This metric could be reported either as an industry-wide aggregate or on a per-company basis. Ideally, improvements in both training compute efficiency and inference compute efficiency should be reported. 

\textbf{Significance}: Compute efficiency improvements on AI R\&D tasks could be a signal of AIRDA. This metric also directly tracks AI progress: if AIRDA increases the pace at which algorithmic or data improvements are created, one expected signal would be large and sustained increases in compute efficiency.

\textbf{Feasibility}: Moderate to high, depending on whether companies need to run new evaluations and train new AI systems. The core methodology involves comparing the performance of models trained or run with various compute budgets \citep{hernandez_measuring_2020,davidson_ai_2023,ho_algorithmic_2024,ho_quantifying_2025,ho_rosetta_2025,whitfill_note_2025,gundlach_origin_2025}. Obtaining accurate estimates of efficiency improvements could require training additional AI systems that companies will not use in production. The cost could potentially be reduced by focusing on improvements in post-training rather than pre-training. Furthermore, incomplete estimates based on existing evaluations and training runs could still be informative. 

\textbf{Limitations}:
\begin{itemize}
    \item \textbf{Limitations of evaluations}: This metric inherits the limits of evaluations discussed for \hyperref[metric:1]{Metric \#1}, such as benchmark saturation and inconsistencies in evaluation set-ups. 
    \item \textbf{Does not capture performance ceiling}: Compute efficiency measures how cheaply a given performance level can be achieved, but it says nothing about the maximum performance attainable. For example, it could become cheaper over time to reach 50\% on a benchmark, while the best achievable score remains stuck at 75\% year over year.
    \item \textbf{Attribution difficulties}: This metric does not distinguish between the causes of observed efficiency gains, which could be due to factors such as algorithmic improvements, improvements in data quality, or scaling up training compute.\footnote{Some algorithmic improvements could be scale-dependent: they provide much larger efficiency gains as compute is scaled up. For example, \citet{gundlach_origin_2025} compare transformers to LSTMs, finding a 6.3x efficiency improvement at $10^{15}$ FLOP and a 26x improvement at $3\cdot 10^{16}$ FLOP.  } If observed gains are due to scaling up training compute, acceleration of AI progress may not necessarily be attributable to AIRDA. 
\end{itemize}

\textbf{Sensitivity}: Low to moderate. Leading companies may report the metric as a positive signal of their research capabilities (e.g., to advertise their products or attract talent), while lagging companies may hesitate to reveal their relative standing. Sensitivity decreases further if reported as an industry-wide aggregate.

\textbf{What remains to be done}: Calculating efficiency improvements from existing data. Potentially running additional evaluations.

\subsection{Survey-Based Metrics}\label{sec:survey}
These metrics involve surveying human researchers at AI companies. 

\subsubsection{Metric \#6: Staff views on AI use and productivity boosts}\label{metric:6}

\metrictable{\cmark}{\cmark}{\cmark}{}{}

\textbf{Description}: This metric captures researchers' self-reported AI use patterns and productivity gains across different task categories. For each task category, researchers answer questions such as:
\begin{itemize}
    \item ``What percentage speed-up do you experience when using AI tools for this task?''
    \item ``Approximately how many hours per week do AI tools save you on this task?''
    \item ``To what extent could AI tools replace a newly hired junior researcher?''
    \item ``How have the tasks in your day-to-day job shifted over the past X months from AI use, if at all?''
    \item ``How much of a speed-up in AI progress would you expect over the next six months as a result of AI tools?''
\end{itemize}
Task categories could include: code implementation and debugging, experiment design and execution, literature review, data analysis, writing and documentation, and review of AI-generated outputs. %Results should be reported as median speed-up multipliers per task category and median hours saved per week, tracked over time.

\textbf{Significance}: Higher productivity boosts could suggest greater AIRDA, and productivity boosts for oversight tasks could suggest that oversight capacity is increasing. Comparing productivity boosts across teams (e.g., safety vs.\ pre-training) could provide information about differential AI progress. Open-ended questions about AI use patterns could also complement more quantitative data.

\textbf{Feasibility}: High. Implementation requires only periodic surveys with straightforward questions. Most organizations already have infrastructure for such data collection. As further evidence of feasibility, the Claude Opus 4.5 System Card includes data from an internal survey that asks whether Opus ``could completely automate a junior ML researcher'' \citep{anthropic_system_2025}. The main challenge is maintaining response rates and avoiding survey fatigue.

\textbf{Limitations}:
\begin{itemize}
    \item \textbf{Subjectivity}: Self-reported productivity gains are subjective and may not correlate with objective output measures. Survey respondents could also overestimate benefits due to novelty effects, or underestimate them as they grow accustomed to AI tools and raise their expectations of what counts as a meaningful boost~\citep{becker_measuring_2025}.
    \item \textbf{Coverage gaps}: Productivity gains from automation of a task could go unreported if researchers no longer consider it part of their job. This metric also does not capture the creation of new kinds of AI R\&D tasks as a result of automation (e.g.\ substantially more time spent reviewing AI logs).
\end{itemize}

\textbf{Sensitivity}: Low. The data is subjective and provides limited competitive intelligence. Some organizations might voluntarily share positive results for recruiting and marketing purposes.

\textbf{What remains to be done}: Designing questions and establishing a regular cadence for running surveys.

\subsubsection{Metric \#7: Extent of AI use in high-stakes decisions}\label{metric:7}

\metrictable{\cmark}{}{}{\cmark}{\cmark}

\textbf{Description}: This metric surveys researchers' judgment of how AI is involved in categories of high-stakes decisions, such as scaling up or launching training runs, compute allocation across teams, and deployment decisions. For each decision, the survey would assess both the level and nature of AI involvement. Potential questions could include:
\begin{itemize}
    \item For this (category of) decision, select all below that apply:
    \begin{itemize}
        \item AI generated initial options/proposals that humans chose from
        \item AI provided analysis or data synthesis that informed the decision
        \item AI drafted decision memos or documentation
        \item AI made recommendations that were adopted without significant modification
        \item AI flagged risks or concerns that changed the decision
    \end{itemize}
    \item Overall, how crucial was the involvement of an AI system in making this decision? (reported on a Likert scale, with optional comment fields)
    \item What kind of oversight was exercised over the AI system? 
\end{itemize}

\textbf{Significance}: Increased AI involvement in high-stakes decisions could be a signal of AIRDA. Increased involvement could also mean increased oversight demand: failure of such AI systems would be more consequential given the stakes of the decisions at hand. This metric can also track how much oversight is performed if the survey asks about it.

\textbf{Feasibility}: Low to moderate. Data collection is relatively straightforward, but companies would need to define which high-stakes decisions to track and how to assess AI involvement in such decisions. One way of collecting the data is to integrate the survey into existing approval workflows for major decisions (e.g., training run sign-offs).

\textbf{Limitations}:
\begin{itemize}
    \item \textbf{Subjectivity}: Assessing the importance of AI involvement would likely be subjective. Researchers could potentially have reputational incentives to understate AI involvement. 
    \item \textbf{Does not capture decision quality}: If decisions are of poor quality when AI involvement is high, AIRDA could be less extensive, or less likely to increase AI progress, than this metric would suggest.
\end{itemize}

\textbf{Sensitivity}: High. This metric likely reveals information about internal decision-making processes and strategy. 

\textbf{What remains to be done}: Identifying high-stakes R\&D decisions. Designing surveys and running them each time such decisions are made.

\subsection{Operational Metrics}\label{sec:operational}
These metrics monitor ongoing R\&D processes and events.

\subsubsection{Metric \#8: Researcher time allocation (``AI-powered Toggl'')}\label{metric:8}

\metrictable{\cmark}{}{}{\cmark}{}

\textbf{Description}: This metric would use software to track how AI researchers allocate their time across different R\&D activities. Such software could use AI systems to categorize the activities on the researcher's screen. 

Task categories could include:
\begin{itemize}
    \item Primary R\&D activities:
    \begin{itemize}
        \item Writing and debugging code
        \item Designing experiments and formulating hypotheses
        \item Running and monitoring experiments
        \item Analyzing experimental results and interpreting data
        \item Literature review and reading research papers
        \item Architecture and system design
        \item Meetings and collaboration
        \item Infrastructure and tooling setup
    \end{itemize}
    \item AI interaction activities:
    \begin{itemize}
        \item Prompting or directing AI coding assistants
        \item Reviewing AI-generated outputs (code, analysis, summaries)
        \item Debugging AI behaviour or outputs
    \end{itemize}
\end{itemize}
Results should be reported as median and mean percentage allocation per activity category across all researchers, tracked over time.

\textbf{Significance}: Time allocation patterns could provide on-the-ground evidence of the extent of AIRDA in a way that benchmarks cannot capture. For example, it would be strong evidence of AIRDA if researchers shifted more of their time from primary R\&D activities to AI interaction activities. Time spent on oversight activities also measures how much oversight is performed.

\textbf{Feasibility}: Moderate. Time-tracking software needs to be written and there needs to be ways to navigate privacy concerns. Tracking will likely become significantly easier over time because AI tools can automate it. 

\textbf{Limitations}:
\begin{itemize}
    \item \textbf{Time spent is not a perfect indicator of productivity or output quality}: For example, code quality is not necessarily correlated with the time spent on coding.  
    \item \textbf{Difficulty aggregating results}: It may be difficult to compare and aggregate time allocation between researchers at different skill levels or working on different problem types.
    \item \textbf{Unobservables}: The metric does not capture cognitive load. Researchers might spend less wall-clock time on a task but more mental effort reviewing AI outputs. It might also be difficult to measure activities that are not captured in software applications (e.g.\ time spent outside of the office thinking about a problem). 
    \item \textbf{Confounders}: Time allocation could shift without corresponding changes in AIRDA, such as a result of shifting company priorities.
\end{itemize}

\textbf{Sensitivity}: Low to moderate, depending on the granularity of the activity categories. Time allocation data reveals workflow patterns and what AI is successfully automating, which could provide some competitive intelligence. Sensitivity is likely to be higher for detailed breakdowns because they could reveal strategic priorities or weaknesses. Time-tracking software itself could also raise privacy concerns amongst employees.

\textbf{What remains to be done}: Developing and deploying the time-tracking software, managing privacy concerns.

\subsubsection{Metric \#9: Oversight effectiveness retrospectives}\label{metric:9}

\metrictable{}{}{}{\cmark}{\cmark}

\textbf{Description}: This metric focuses on defects in AI-generated R\&D outputs. For each AI-generated R\&D output, companies would track:
\begin{itemize}
    \item Whether the output underwent human review (and what kind)
    \item The degree to which AI systems assisted with human review
    \item A description of the defects, including bugs, errors, security vulnerabilities, misconfigurations, or other quality issues
    \item When the defects were discovered (e.g., during review, after review but before production, or after production)
    \item The severity of the defects discovered
\end{itemize}

Categories of AI R\&D outputs could include randomly selected subsets of:
\begin{itemize}
    \item Pull requests
    \item Analyses of experimental data
    \item Logs of how a training run was monitored
\end{itemize}

Both the proportion of outputs that have defects and the length of time until discovery of the defect(s) should be reported, stratified by severity.

\textbf{Significance}: This metric directly tracks how much oversight is performed on AI-generated outputs. Increases in the frequency of defects would suggest increases in the oversight gap. If the level of oversight is held constant, increases in the frequency of defects could also suggest increases in oversight demand. 

\textbf{Feasibility}: Low to moderate, depending on the type of output. One difficulty is defining a defect. For code, defining defects is relatively straightforward. For scientific papers, reasonable people may disagree on what counts as a defect, and such defects may only come to light long after publication. Another difficulty is that in practice, outputs will be partly human-written and partly AI-generated. Organizations may need to build systems to track which portions are AI-generated. For example, \citet{cursor_agent_2026} proposes a standard for labelling AI-generated code.

\textbf{Limitations}:
\begin{itemize}
    \item \textbf{Limited comparability}: There can be reasonable disagreement about what constitutes a defect in a complicated output such as a scientific paper. Inconsistent definition of defects could make it difficult to compare results between AI companies. Inconsistencies in how human review is defined between companies (e.g.\ quick glance vs.\ detailed audit) could also hinder comparability.
    \item \textbf{Difficulty of catching defects}: Defect detection depends on downstream quality processes (testing, review, monitoring), which may not catch all errors equally for AI versus human work (e.g.\ current error detection processes might be more effective at catching human errors). 
\end{itemize}

\textbf{Sensitivity}: Moderate to high because the metric could reveal oversight quality problems and raise safety concerns.

\textbf{What remains to be done}: Building a tracking/tagging pipeline and establishing a regular data analysis schedule.

\subsubsection{Metric \#10: AI subversion incidents}\label{metric:10}

\metrictable{}{}{}{}{\cmark}

\textbf{Description}: This metric tracks incidents where AI systems have attempted to subvert AI R\&D workflows in practice. Examples of subversive actions could include sabotage of experiments, insertion of backdoors into production code, or manipulation of evaluation results. For each incident, organizations would record:
\begin{itemize}
    \item A description of the subversive action attempted
    \item When the incident occurred
    \item The context in which it occurred (e.g., which task, which model, what level of autonomy)
    \item How the incident was detected (e.g., chain-of-thought analysis)
    \item The severity of the incident (i.e., the damage caused or the potential damage had the attempt succeeded)
    \item An assessment of why the incident occurred, including the extent to which the incident appears to be goal-directed, as opposed to being a result of e.g. misunderstanding instructions. 
\end{itemize}
As a summary statistic, the number and severity of such subversion incidents could be reported. 

\textbf{Significance}: An increasing number or severity of subversion attempts, even if detected and prevented, indicates increased oversight demand. Compared to \hyperref[metric:9]{Metric \#9}, which measures defects in AI-generated outputs, this metric focuses on subversion attempts, whether successful or not. Compared to \hyperref[metric:3]{Metrics \#3} and \hyperref[metric:4]{\#4}, this metric focuses on real production workflows rather than sandboxed or toy environments.

\textbf{Feasibility}: Low to moderate. The monitoring infrastructure to detect subversion could require significant engineering effort and compute. Some companies are beginning to build such infrastructure \citep{openai_gpt53codex_2026}.

\textbf{Limitations}:
\begin{itemize}
    \item \textbf{Definitional ambiguity}: It may be difficult to define what counts as a subversive action. Apparently subversive actions could result from factors such as capability limitations or ambiguous instructions. Models can also produce misleading or post-hoc rationalizations in their reasoning traces.
    \item \textbf{Under-detection}: Because AI systems could try to evade detection, the observed incident count could significantly underestimate the true rate.
\end{itemize}

\textbf{Sensitivity}: Moderate to high. Public knowledge of subversion attempts could undermine confidence in the company's products and reveal specific vulnerabilities.

\textbf{What remains to be done}: Defining what counts as a subversion incident and building detection infrastructure.

\subsection{Organizational Metrics}\label{sec:organizational}
These metrics capture organizational structure, resource allocation, and policies.

\subsubsection{Metric \#11: Headcount of AI researchers and the distribution of their performance}\label{metric:11}

\metrictable{\cmark}{\cmark}{\cmark}{\cmark}{}

\textbf{Description}: This metric measures the number and seniority of AI researchers at frontier AI companies, along with the distribution of their performance. Headcount should be disaggregated by team (e.g.\ trust and safety team, pre-training team, etc). Performance could be measured (or proxied) in a variety of ways, such as:
\begin{itemize}
    \item Completion of OKRs (adjusted by difficulty)
    \item Number of code commits
    \item Salaries
    \item Compute budgets
\end{itemize}

Researcher headcount should not include:
\begin{itemize}
    \item Staff building software products, features, or services on top of existing AI systems
    \item Business, operational, marketing, legal staff, etc.
\end{itemize}

\textbf{Significance}: AIRDA could lead to decreases in headcount if a small number of researchers is able to get much of the necessary R\&D work done with AI assistance. As explained in \Cref{sec:oversight}, lower headcounts could mean reduced oversight capacity and level of oversight achieved. Furthermore, a widening distribution of researcher performance could signal that AI tools are becoming powerful enough to create large productivity differences between those who use them effectively and those who do not. As AI systems continue to improve, this gap could compound.

\textbf{Feasibility}: Moderate to high. HR already maintains records of staff counts by role type and salary levels. Compiling and reporting this information requires minimal additional infrastructure. However, performance data may be more difficult to standardize within and across organizations.

\textbf{Limitations}:
\begin{itemize}
    \item \textbf{Difficulties measuring performance}: Performance metrics like OKRs may not capture the most important research contributions, which are often difficult to predict or measure. Attribution of research outcomes to individual researchers is inherently difficult in collaborative environments. Research is also non-linear: long periods of apparent unproductivity could precede major breakthroughs. 
    \item \textbf{Confounders for headcount}: Headcount changes could reflect a variety of other factors such as funding cycles, labour market conditions, or strategic pivots.
    \item \textbf{Non-linearity}: We should expect human labour to become more valuable before full automation because AI systems can augment human labour~\citep{gans_oring_2026}. As a result, headcount could remain high or even increase as long as human-dependent activities bottleneck full automation, such as providing high-level research direction \citep{toner_when_2026}.
\end{itemize}

\textbf{Sensitivity}: Moderate to high. Sharing salaries in particular could invite public criticism or create uncomfortable social dynamics between employees. Sharing OKRs could reveal research strategies. However, some information (headcount, hiring trends) can be gathered through publicly available sources such as LinkedIn. 

\textbf{What remains to be done}: Aggregating and presenting existing data.

\subsubsection{Metric \#12: Distribution of compute usage}\label{metric:12}

\metrictable{\cmark}{}{}{}{}

\textbf{Description}: This metric measures the amount of compute (FLOP) used by frontier AI companies across the AI R\&D process, including pre-training, mid-training, post-training, evaluation, experimentation, internal inference (e.g.\ coding and research agents), and external deployment.

\textbf{Significance}: Shifts in compute allocation towards internal inference, especially relative to compute used for external deployment, could suggest increases in AIRDA. Comparing internal and external inference could help to distinguish AIRDA from company growth. 

\textbf{Feasibility}: Moderate to high. Compute is likely to be meticulously tracked at frontier AI companies for cost management and infrastructure planning. For example, pre-training workloads require months-long planning and dedicated hardware reservations. The main implementation challenge is categorizing uses of compute. Public estimates based on dollar spend are possible~\citep{you_most_2025}, but would not track fine-grained details about how compute is used.

\textbf{Limitations}:
\begin{itemize}
    \item \textbf{Other factors affecting compute allocation}: Breakthroughs in compute efficiency could mean that less compute is required to accomplish the same activity. The allocation between external and internal inference could also heavily depend upon market dynamics and investor demands. 
    \item \textbf{Inconsistent categorization of activities}: Because different AI companies could have different R\&D workflows, results may not be comparable between them.
\end{itemize}

\textbf{Sensitivity}: High. Compute usage data reveals both absolute resource levels relative to competitors and R\&D productivity. Many frontier AI companies stopped disclosing pre-training compute in 2024, although the EU AI Act~\citep{regulation_2024} requires some disclosure for in-scope systems. 

\textbf{What remains to be done}: Developing systems to classify compute use. For example, a workload classifier could tag jobs submitted to a compute cluster, while another classifier could tag internal agent or chat logs, similar to \citet{tamkin_clio_2024}. External organizations can try to create estimates of this metric based on publicly available information.

\subsubsection{Metric \#13: Capital share of AI R\&D spending}\label{metric:13}

\metrictable{\cmark}{}{}{}{}

\textbf{Description}: This metric measures compute expenditure as a percentage of total AI R\&D spending.\footnote{A broader variant of this metric is the network-adjusted capital share \citep{trammell_what_2026}, which would measure the extent to which a particular process (e.g. AI R\&D) could sustain itself without human labour anywhere in its supply chain.} It is calculated as: 

\begin{equation*}
\frac{\text{compute costs for R\&D activities}}{\text{total R\&D expenditure including labour, compute, and other costs}} \times 100.    
\end{equation*}

\textbf{Significance}: As AI systems take on more of the work previously done by humans, the share of spending going to compute (capital) should increase relative to spending on labour and other costs. Combining this metric with \hyperref[metric:12]{Metric \#12} could help to distinguish increased spending on inference from increased spending on training.

\textbf{Feasibility}: High. Organizations already track R\&D expenditures for internal budgeting and tax purposes, and standardized tax definitions of R\&D could facilitate cross-company comparisons. Separating compute costs from labour costs is straightforward accounting.

\textbf{Limitations}:
\begin{itemize}
    \item \textbf{Ambiguity}: Rising capital share could reflect increasing AIRDA, but could also reflect larger investments in training runs, changes in compute pricing (e.g., from supply chain disruptions), or wasteful or inefficient uses of compute.
\end{itemize}

\textbf{Sensitivity}: Moderate. This metric reveals the basic structure of R\&D spending, but it does not directly expose capabilities, specific techniques, or detailed competitive information.

\textbf{What remains to be done}: Calculating and reporting the metric based on existing data.

\subsubsection{Metric \#14: AI permission lists}\label{metric:14}

\metrictable{\cmark}{}{}{}{\cmark}

\textbf{Description}: A list of which actions AI systems are authorized to take and what level of human review or approval is needed, if any. Actions could include: modifying production code, initiating training runs, deploying models, and accessing sensitive systems or data. For each action, organizations would report the default permission level (e.g.\ no review required), recording notable deviations if applicable. 

\textbf{Significance}: This metric tracks oversight demand, which could be higher if AI systems are authorized to take significant actions without human review. AIRDA is likely also higher if AI systems are permitted to take significant actions autonomously (e.g., launching large training runs). Compared to \hyperref[metric:7]{Metric \#7}, which is about AI use in practice, this metric is about policies around AI use.

\textbf{Feasibility}: Moderate to high. Organizations implementing AI agents for R\&D tasks must already make decisions about permissions and access controls. This metric simply requires documenting those decisions systematically.

\textbf{Limitations}:
\begin{itemize}
    \item \textbf{Formal vs.\ actual permissions}: Documented permissions may not reflect actual practice. For example, exceptions to the rules might be granted easily, or human approval could be perfunctory.
    \item \textbf{Granularity challenges}: Actions must be specific enough to be meaningful, but ideally general enough to be comparable across different organizations with different infrastructure.
    \item \textbf{Does not account for reliability}: Even if AI systems are authorized to take high-stakes actions, oversight demand could in principle be lower if AI systems are much more reliable than humans. 
\end{itemize}

\textbf{Sensitivity}: Moderate to high, depending on the level of granularity. Permission lists could reveal internal security practices and potential vulnerabilities. High-level summaries (e.g., ``AI systems cannot autonomously modify production code'' vs.\ ``AI systems can autonomously commit to any repository'') would likely be less sensitive.

\textbf{What remains to be done}: Defining actions of interest, collecting permissions across teams, and updating the list regularly.

\section{Limitations}\label{sec:limitations}
This work has several limitations that warrant consideration.

Our analysis of AIRDA's implications in \Cref{sec:implications} has largely ignored higher-order effects. While their significance is highly uncertain, they could be meaningful to track. For example, if AI R\&D automation advances safety research faster than capabilities research, the perception that safety is progressing adequately could encourage more risk-taking. 

Many of the proposed metrics are lagging indicators: they measure automation that has already occurred. By the time these metrics show concerning trends, the window for potential interventions may have narrowed considerably. While benchmark performance on AI R\&D tasks could be a leading indicator, translating such performance into predictions about real-world AIRDA remains imprecise.

Relatedly, the relationship between AIRDA metrics and AI progress could be highly non-linear. AI R\&D could be substantially automated across most tasks while remaining bottlenecked by a single human-dependent activity, such as providing high-level research direction. In such scenarios, metrics might indicate high levels of automation without corresponding acceleration of AI progress. Automation of that final bottleneck could dramatically accelerate progress, yet only show up as a small change in the metrics.

Differences in data collection methods could limit comparability between AI companies. For example, companies could define task categories differently or use different attribution methods for AI contributions. Without standardized measurement protocols, aggregating or comparing metrics across organizations could be misleading. This limitation is particularly acute for metrics involving subjective judgment, such as surveys on productivity gains.

Given that AI companies would internally collect most of our proposed metrics, it might be difficult to verify their accuracy. For instance, companies could understate the degree of automation for fear of government intervention. As a mitigation, independent third parties could help to corroborate company results, as they already do for benchmarks \citep{ukaisecurityinstitute_predeployment_2024,schoen_stress_2025,epochai_introducing_2024}. AI companies could also contract external organizations to carry out data collection with standardized protocols, such as for researcher surveys. 

Any collection of AIRDA metrics will likely need to be updated over time. In addition to lessons learned over the course of developing and collecting such metrics, we will likely gain more clarity over time about which effects of AIRDA to prioritize. 

Finally, many factors relevant to AI progress and oversight fall outside the domain of AI R\&D. For instance, metrics focused on AIRDA do not capture the cybersecurity readiness of critical infrastructure providers \citep{kamilelukosiute_design_2025} or the pace of regulatory adaptation. AI companies would also find it difficult to track such metrics by themselves. A comprehensive assessment of AIRDA's implications will likely require cross-domain collaborations.

\section{Related Work}
A growing suite of benchmarks evaluates AI systems on tasks relevant to AI R\&D. SWE-bench and its variants \citep{jimenez_swebench_2023} assess software engineering capabilities on real GitHub issues, while MLE-bench evaluates ML engineering through Kaggle competitions \citep{chan_mlebench_2025}. For tasks more directly relevant to frontier research, METR's RE-Bench compares AI agents to human experts on ML research engineering tasks \citep{wijk_rebench_2025}, and OpenAI's PaperBench tests agents' ability to replicate ICML papers from scratch \citep{starace_paperbench_2025}. 

A number of works have studied theoretical models of AI R\&D automation. \citet{eth_will_2025} explore the possibility that software progress alone could lead to increasingly accelerating AI progress. \citet{aifuturesproject_ai_2026} provides a model for forecasting AI R\&D automation. \citet{erdil_most_2025} argue that automating AI R\&D alone likely will not accelerate AI progress. \citet{ho_software_2025} point out problems both with existing data sources and the models used to study the effects of AI R\&D automation. \citet{erdil_gate_2025} model the feedback loop that AI development could create between itself and the economy. \citet{whitfill_will_2025} argue that compute and cognitive labour are complements in the context of frontier experiments.  

Under both voluntary commitments and legislation, AI companies will likely report some details about AI R\&D automation. Many safety frameworks address AI R\&D automation. Anthropic's Responsible Scaling Policy \citep{anthropic_responsible_2026} defines a threshold for AI R\&D automation based on compressing ``two years of 2018 – 2024 AI progress into a single year'' . OpenAI's Preparedness Framework \citep{openai_preparedness_2025} tracks ``AI Self-improvement'' capabilities, with a ``High'' threshold for models whose impact is ``equivalent to giving every OpenAI researcher a highly performant mid-career research engineer assistant, relative to those researchers’ 2024 baseline''. Google DeepMind's Frontier Safety Framework \citep{googledeepmind_frontier_2025} includes ``Machine Learning R\&D'' as a Critical Capability Level domain, recommending high security measures for models that could significantly accelerate AI development. Furthermore, under California's Transparency in Frontier Artificial Intelligence Act, each large frontier developer must describe how it approaches ``[a]ssessing and managing catastrophic risk resulting from the internal use of its frontier models, including risks resulting from a frontier model circumventing oversight mechanisms.'' Some recent work has begun to study how internal use of automated AI researchers could lead to such risks \citep{clymer_bare_2025,benton_sabotage_2024,stix_ai_2025,metr_ai_2025,korbak_sketch_2025,greenblatt_prioritizing_2024}.

\citet{toner_when_2026} offers a useful complement to this piece, distilling areas of expert consensus and disagreement from a workshop on AI R\&D automation. 

\section{Conclusion}
AI R\&D automation could have significant effects on AI progress and human oversight. We have proposed a variety of metrics to help track and understand these effects. Companies, governments, and third parties (e.g. non-profit research organisations) all have roles in advancing these metrics. We especially recommend prioritizing metrics (or components thereof) that are relatively neglected (\Cref{tab:recommendations}).

In addition to clarifying the degree of AIRDA and its implications, future work should also investigate how to respond to different AIRDA scenarios. Promising areas of investigation could include how to accelerate defensive AI capabilities, how to improve the ability of institutions to adapt to rapid progress, and whether and how to impose human oversight requirements.

\subsubsection*{Acknowledgments}
The following people participated in a preliminary workshop and survey that informed this work: Ardi Janjeva, Annika Hallensleben, Dewey Murdick, Vasilios Mavroudis, Helen Toner, Eli Lifland, Carolyn Ashurst, David Owen, Ture Hinrichsen, Rosco Hunter, and other participants who prefer to remain anonymous. We would also like to thank the following people for useful feedback on previous drafts of this work: Tom Davidson, Sam Manning, Parker Whitfill, Cheryl Wu, Micah Carroll, Hjalmar Wijk, and Anson Ho. All errors remain our own. 

\bibliography{main_clean}

@misc{trammell_what_2026,
	type = {Substack newsletter},
	title = {What (and how far off) is self-replicating capital?},
	url = {https://philiptrammell.substack.com/p/what-and-how-far-off-is-self-replicating},
	urldate = {2026-03-04},
	journal = {Philosopher count},
	author = {Trammell, Philip},
	month = jan,
	year = {2026},
}

@misc{alphabet_alphabet_2026,
	title = {Alphabet {Q4} 2025 earnings call},
	url = {https://abc.xyz/investor/events/event-details/2026/2025-Q4-Earnings-Call-2026-Dr_C033hS6/default.aspx},
	author = {{Alphabet}},
	year = {2026},
}

@misc{whitfill_note_2025,
	title = {Note on {Selection} {Bias} in {Observational} {Estimates} of {Algorithmic} {Progress}},
	url = {http://arxiv.org/abs/2508.11033},
	doi = {10.48550/arXiv.2508.11033},
	urldate = {2026-02-28},
	publisher = {arXiv},
	author = {Whitfill, Parker},
	month = aug,
	year = {2025},
	note = {arXiv:2508.11033 [econ]},
}

@misc{gundlach_origin_2025,
	title = {On the {Origin} of {Algorithmic} {Progress} in {AI}},
	url = {http://arxiv.org/abs/2511.21622},
	doi = {10.48550/arXiv.2511.21622},
	urldate = {2026-02-28},
	publisher = {arXiv},
	author = {Gundlach, Hans and Fogelson, Alex and Lynch, Jayson and Trisovic, Ana and Rosenfeld, Jonathan and Sandhu, Anmol and Thompson, Neil},
	month = nov,
	year = {2025},
	note = {arXiv:2511.21622 [cs]},
}

@misc{hernandez_measuring_2020,
	title = {Measuring the {Algorithmic} {Efficiency} of {Neural} {Networks}},
	url = {http://arxiv.org/abs/2005.04305},
	doi = {10.48550/arXiv.2005.04305},
	urldate = {2026-02-28},
	publisher = {arXiv},
	author = {Hernandez, Danny and Brown, Tom B.},
	month = may,
	year = {2020},
	note = {arXiv:2005.04305 [cs]},
}

@misc{davidson_ai_2023,
	title = {{AI} capabilities can be significantly improved without expensive retraining},
	url = {http://arxiv.org/abs/2312.07413},
	doi = {10.48550/arXiv.2312.07413},
	urldate = {2026-02-28},
	publisher = {arXiv},
	author = {Davidson, Tom and Denain, Jean-Stanislas and Villalobos, Pablo and Bas, Guillem},
	month = dec,
	year = {2023},
	note = {arXiv:2312.07413 [cs]},
}

@misc{ho_quantifying_2025,
	title = {Quantifying the algorithmic improvement from reasoning models},
	url = {https://epoch.ai/gradient-updates/quantifying-the-algorithmic-improvement-from-reasoning-models},
	author = {Ho, Anson and Berg, Arden},
	year = {2025},
}

@misc{ho_rosetta_2025,
	title = {A {Rosetta} {Stone} for {AI} {Benchmarks}},
	url = {https://epoch.ai/blog/a-rosetta-stone-for-ai-benchmarks},
	author = {Ho, Anson and Denain, Jean-Stanislas and Atanasov, David and Albanie, Samuel and Shah, Rohin},
	year = {2025},
}

@misc{ho_least_2026,
	title = {The least understood driver of {AI} progress},
	url = {https://epoch.ai/gradient-updates/the-least-understood-driver-of-ai-progress},
	author = {Ho, Anson},
	year = {2026},
}

@techreport{anthropic_responsible_2026,
	title = {Responsible {Scaling} {Policy}},
	number = {Version 3},
	author = {{Anthropic}},
	year = {2026},
}

@misc{becker_we_2026,
	title = {We are changing our developer productivity experiment design},
	url = {https://metr.org/blog/2026-02-24-uplift-update/},
	author = {Becker, Joel and Rush, Nate and Cunningham, Tom and Rein, David and Mahamud, Khalid},
	month = feb,
	year = {2026},
}

@misc{brundage_frontier_2026,
	title = {Frontier {AI} {Auditing}: {Toward} {Rigorous} {Third}-{Party} {Assessment} of {Safety} and {Security} {Practices} at {Leading} {AI} {Companies}},
	shorttitle = {Frontier {AI} {Auditing}},
	url = {http://arxiv.org/abs/2601.11699},
	doi = {10.48550/arXiv.2601.11699},
	urldate = {2026-02-24},
	publisher = {arXiv},
	author = {Brundage, Miles and Dreksler, Noemi and Homewood, Aidan and McGregor, Sean and Paskov, Patricia and Stosz, Conrad and Sastry, Girish and Cooper, A. Feder and Balston, George and Adler, Steven and Casper, Stephen and Anderljung, Markus and Werner, Grace and Mindermann, Soren and Mavroudis, Vasilios and Bucknall, Ben and Stix, Charlotte and Freund, Jonas and Pacchiardi, Lorenzo and Hernandez-Orallo, Jose and Pistillo, Matteo and Chen, Michael and Painter, Chris and Ball, Dean W. and O'Keefe, Cullen and Weil, Gabriel and Harack, Ben and Finley, Graeme and Hassan, Ryan and Emmons, Scott and Foster, Charles and Reuel, Anka and Treece, Bri and Bengio, Yoshua and Reti, Daniel and Bommasani, Rishi and Trout, Cristian and Shamsabadi, Ali Shahin and Dattani, Rajiv and Weller, Adrian and Trager, Robert and Sevilla, Jaime and Wagner, Lauren and Soder, Lisa and Ramakrishnan, Ketan and Papadatos, Henry and Murray, Malcolm and Tovcimak, Ryan},
	month = feb,
	year = {2026},
	note = {arXiv:2601.11699 [cs]},
}

@techreport{openai_gpt53codex_2026,
	title = {{GPT}-5.3-{Codex} {System} {Card}},
	author = {{OpenAI}},
	month = feb,
	year = {2026},
}

@misc{cursor_agent_2026,
	title = {Agent {Trace}},
	url = {https://agent-trace.dev/},
	language = {en},
	urldate = {2026-02-19},
	author = {{Cursor}},
	year = {2026},
}

@misc{marks_auditing_2025,
	title = {Auditing language models for hidden objectives},
	url = {http://arxiv.org/abs/2503.10965},
	doi = {10.48550/arXiv.2503.10965},
	urldate = {2026-02-19},
	publisher = {arXiv},
	author = {Marks, Samuel and Treutlein, Johannes and Bricken, Trenton and Lindsey, Jack and Marcus, Jonathan and Mishra-Sharma, Siddharth and Ziegler, Daniel and Ameisen, Emmanuel and Batson, Joshua and Belonax, Tim and Bowman, Samuel R. and Carter, Shan and Chen, Brian and Cunningham, Hoagy and Denison, Carson and Dietz, Florian and Golechha, Satvik and Khan, Akbir and Kirchner, Jan and Leike, Jan and Meek, Austin and Nishimura-Gasparian, Kei and Ong, Euan and Olah, Christopher and Pearce, Adam and Roger, Fabien and Salle, Jeanne and Shih, Andy and Tong, Meg and Thomas, Drake and Rivoire, Kelley and Jermyn, Adam and MacDiarmid, Monte and Henighan, Tom and Hubinger, Evan},
	month = mar,
	year = {2025},
	note = {arXiv:2503.10965 [cs]},
}

@misc{regulation_2024,
	title = {Regulation ({EU}) 2024/1689 of the {European} {Parliament} and of the {Council} of 13 {June} 2024 laying down harmonised rules on artificial intelligence and amending {Regulations} ({EC}) {No} 300/2008, ({EU}) {No} 167/2013, ({EU}) {No} 168/2013, ({EU}) 2018/858, ({EU}) 2018/1139 and ({EU}) 2019/2144 and {Directives} 2014/90/{EU}, ({EU}) 2016/797 and ({EU}) 2020/1828 ({Artificial} {Intelligence} {Act})},
	url = {http://data.europa.eu/eli/reg/2024/1689/oj},
	language = {multiple},
	month = jul,
	year = {2024},
	note = {Volume: L
tex.shorthand: EU AI Act},
	pages = {2024/1689},
}

@book{sagan_spread_2003,
	edition = {2nd},
	title = {The spread of nuclear weapons : a debate renewed with new sections on {India} and {Pakistan}, terrorism, and missile defense},
	url = {https://cir.nii.ac.jp/crid/1970586434801795733},
	publisher = {W.W. Norton},
	author = {Sagan, Scott Douglas and Waltz, Kenneth Neal},
	year = {2003},
}

@article{waltz_spread_1981,
	title = {The {Spread} of {Nuclear} {Weapons}: {More} {May} {Be} {Better}: {Introduction}},
	volume = {21},
	issn = {0567-932X},
	shorttitle = {The {Spread} of {Nuclear} {Weapons}},
	url = {https://doi.org/10.1080/05679328108457394},
	doi = {10.1080/05679328108457394},
	number = {171},
	urldate = {2026-02-18},
	journal = {The Adelphi Papers},
	publisher = {Routledge},
	author = {Waltz, Kenneth N.},
	month = sep,
	year = {1981},
	note = {\_eprint: https://doi.org/10.1080/05679328108457394},
	pages = {1--1},
}

@misc{korinek_scenarios_2024,
	type = {Working {Paper}},
	series = {Working {Paper} {Series}},
	title = {Scenarios for the {Transition} to {AGI}},
	url = {https://www.nber.org/papers/w32255},
	doi = {10.3386/w32255},
	urldate = {2026-02-11},
	publisher = {National Bureau of Economic Research},
	author = {Korinek, Anton and Suh, Donghyun},
	month = mar,
	year = {2024},
	doi = {10.3386/w32255},
}

@techreport{anthropic_system_2026,
	title = {System {Card}: {Claude} {Opus} 4.6},
	author = {{Anthropic}},
	month = feb,
	year = {2026},
}

@techreport{xai_xai_2025,
	title = {{xAI} {Risk} {Management} {Framework}},
	author = {{xAI}},
	year = {2025},
}

@misc{whitfill_will_2025,
	title = {Will {Compute} {Bottlenecks} {Prevent} an {Intelligence} {Explosion}?},
	url = {http://arxiv.org/abs/2507.23181},
	doi = {10.48550/arXiv.2507.23181},
	urldate = {2026-01-28},
	publisher = {arXiv},
	author = {Whitfill, Parker and Wu, Cheryl},
	month = aug,
	year = {2025},
	note = {arXiv:2507.23181 [econ]},
}

@article{eth_will_2025,
	title = {Will {AI} {R}\&{D} automation cause a software intelligence explosion?},
	url = {https://www.forethought.org/research/will-ai-r-and-d-automation-cause-a-software-intelligence-explosion},
	author = {Eth, Daniel and Davidson, Tom},
	year = {2025},
}

@misc{ho_software_2025,
	title = {The software intelligence explosion debate needs experiments},
	url = {https://epoch.ai/gradient-updates/the-software-intelligence-explosion-debate-needs-experiments},
	author = {Ho, Anson and Whitfill, Parker},
	year = {2025},
}

@article{jimenez_swebench_2023,
	title = {Swe-bench: {Can} language models resolve real-world github issues?},
	journal = {arXiv preprint arXiv:2310.06770},
	author = {Jimenez, Carlos E and Yang, John and Wettig, Alexander and Yao, Shunyu and Pei, Kexin and Press, Ofir and Narasimhan, Karthik},
	year = {2023},
}

@misc{bernardi_societal_2025,
	title = {Societal {Adaptation} to {Advanced} {AI}},
	url = {http://arxiv.org/abs/2405.10295},
	doi = {10.48550/arXiv.2405.10295},
	urldate = {2026-01-28},
	publisher = {arXiv},
	author = {Bernardi, Jamie and Mukobi, Gabriel and Greaves, Hilary and Heim, Lennart and Anderljung, Markus},
	month = jan,
	year = {2025},
	note = {arXiv:2405.10295 [cs]},
}

@techreport{toner_when_2026,
	title = {When {AI} builds {AI}: {Findings} from a {Workshop} on {Automation} of {AI} {R}\&{D}},
	doi = {10.51593/20250027},
	institution = {Center for Security and Emerging Technology},
	author = {Toner, Helen and Beers, Kendrea and Newman, Steve and Khan, Saif M. and Shea-Blymyer, Colin and Yee, Evelyn and Acharya, Ashwin and Fisher, Kathleen and Scholl, Keller and Wildeford, Peter and Greenblatt, Ryan and Albanie, Samuel and Ballard, Stephanie and Larsen, Thomas},
	month = jan,
	year = {2026},
}

@misc{amodei_adolescence_2026,
	title = {The {Adolescence} of {Technology}},
	url = {https://www.darioamodei.com/essay/the-adolescence-of-technology},
	language = {en},
	urldate = {2026-02-11},
	author = {Amodei, Dario},
	year = {2026},
}

@techreport{anthropic_system_2025,
	title = {System {Card}: {Claude} {Opus} 4.5},
	author = {{Anthropic}},
	month = nov,
	year = {2025},
}

@misc{schoen_stress_2025,
	title = {Stress {Testing} {Deliberative} {Alignment} for {Anti}-{Scheming} {Training}},
	url = {http://arxiv.org/abs/2509.15541},
	doi = {10.48550/arXiv.2509.15541},
	urldate = {2026-01-28},
	publisher = {arXiv},
	author = {Schoen, Bronson and Nitishinskaya, Evgenia and Balesni, Mikita and Højmark, Axel and Hofstätter, Felix and Scheurer, Jérémy and Meinke, Alexander and Wolfe, Jason and Weij, Teun van der and Lloyd, Alex and Goldowsky-Dill, Nicholas and Fan, Angela and Matveiakin, Andrei and Shah, Rusheb and Williams, Marcus and Glaese, Amelia and Barak, Boaz and Zaremba, Wojciech and Hobbhahn, Marius},
	month = sep,
	year = {2025},
	note = {arXiv:2509.15541 [cs]},
}

@misc{choi_scaling_2024,
	title = {Scaling automatic neuron description},
	url = {https://transluce.org/neuron-descriptions},
	author = {Choi, Dami and Huang, Vincent and Meng, Kevin and Johnson, Daniel D and Steinhardt, Jacob and Schwettmann, Sarah},
	month = oct,
	year = {2024},
}

@misc{bellan_sam_2025,
	title = {Sam {Altman} says {OpenAI} will have a ‘legitimate {AI} researcher’ by 2028},
	url = {https://techcrunch.com/2025/10/28/sam-altman-says-openai-will-have-a-legitimate-ai-researcher-by-2028/},
	language = {en-US},
	urldate = {2025-12-17},
	journal = {TechCrunch},
	author = {Bellan, Rebecca},
	month = oct,
	year = {2025},
}

@misc{benton_sabotage_2024,
	title = {Sabotage {Evaluations} for {Frontier} {Models}},
	url = {http://arxiv.org/abs/2410.21514},
	doi = {10.48550/arXiv.2410.21514},
	urldate = {2026-01-28},
	publisher = {arXiv},
	author = {Benton, Joe and Wagner, Misha and Christiansen, Eric and Anil, Cem and Perez, Ethan and Srivastav, Jai and Durmus, Esin and Ganguli, Deep and Kravec, Shauna and Shlegeris, Buck and Kaplan, Jared and Karnofsky, Holden and Hubinger, Evan and Grosse, Roger and Bowman, Samuel R. and Duvenaud, David},
	month = oct,
	year = {2024},
	note = {arXiv:2410.21514 [cs]},
}

@misc{greenblatt_prioritizing_2024,
	title = {Prioritizing threats for {AI} control},
	url = {https://blog.redwoodresearch.org/p/prioritizing-threats-for-ai-control},
	language = {en},
	urldate = {2026-01-28},
	author = {Greenblatt, Ryan},
	month = jun,
	year = {2024},
}

@article{macaskill_preparing_2025,
	title = {Preparing for the intelligence explosion},
	url = {https://www.forethought.org/research/preparing-for-the-intelligence-explosion},
	author = {MacAskill, William and Moorhouse, Fin},
	year = {2025},
}

@techreport{openai_preparedness_2025,
	title = {Preparedness {Framework}},
	number = {Version 2},
	author = {{OpenAI}},
	month = apr,
	year = {2025},
}

@misc{ukaisecurityinstitute_predeployment_2024,
	title = {Pre-{Deployment} evaluation of {OpenAI}’s o1 model {\textbar} {AISI} {Work}},
	url = {https://www.aisi.gov.uk/blog/pre-deployment-evaluation-of-openais-o1-model},
	language = {en},
	urldate = {2026-01-28},
	journal = {AI Security Institute},
	author = {{UK AI Security Institute}},
	year = {2024},
}

@techreport{anthropic_responsible_2025,
	title = {Responsible {Scaling} {Policy}},
	number = {Version 2.2},
	author = {{Anthropic}},
	year = {2025},
}

@misc{wijk_rebench_2025,
	title = {{RE}-{Bench}: {Evaluating} frontier {AI} {R}\&{D} capabilities of language model agents against human experts},
	shorttitle = {{RE}-{Bench}},
	url = {http://arxiv.org/abs/2411.15114},
	doi = {10.48550/arXiv.2411.15114},
	urldate = {2026-01-28},
	publisher = {arXiv},
	author = {Wijk, Hjalmar and Lin, Tao and Becker, Joel and Jawhar, Sami and Parikh, Neev and Broadley, Thomas and Chan, Lawrence and Chen, Michael and Clymer, Josh and Dhyani, Jai and Ericheva, Elena and Garcia, Katharyn and Goodrich, Brian and Jurkovic, Nikola and Karnofsky, Holden and Kinniment, Megan and Lajko, Aron and Nix, Seraphina and Sato, Lucas and Saunders, William and Taran, Maksym and West, Ben and Barnes, Elizabeth},
	month = may,
	year = {2025},
	note = {arXiv:2411.15114 [cs]},
}

@misc{wen_predicting_2025,
	title = {Predicting {Empirical} {AI} {Research} {Outcomes} with {Language} {Models}},
	url = {http://arxiv.org/abs/2506.00794},
	doi = {10.48550/arXiv.2506.00794},
	urldate = {2026-01-28},
	publisher = {arXiv},
	author = {Wen, Jiaxin and Si, Chenglei and Chen, Yueh-han and He, He and Feng, Shi},
	month = jun,
	year = {2025},
	note = {arXiv:2506.00794 [cs]
version: 1},
}

@techreport{bengio_international_2026,
	title = {International {AI} {Safety} {Report} 2026},
	url = {https://internationalaisafetyreport.org},
	number = {DSIT 2026/001},
	institution = {Department for Science, Innovation and Technology},
	author = {Bengio, Y. and Clare, S. and Prunkl, C. and Murray, M. and Andriushchenko, M. and Bucknall, B. and Bommasani, R. and Casper, S. and Davidson, T. and Douglas, R. and Duvenaud, D. and Fox, P. and Gohar, U. and Hadshar, R. and Ho, A. and Hu, T. and Jones, C. and Kapoor, S. and Kasirzadeh, A. and Manning, S. and Maslej, N. and Mavroudis, V. and McGlynn, C. and Moulange, R. and Newman, J. and Ng, K. Y. and Paskov, P. and Rismani, S. and Sastry, G. and Seger, E. and Singer, S. and Stix, C. and Velasco, L. and Wheeler, N. and Acemoglu, D. and Conitzer, V. and Dietterich, T. G. and Felten, E. W. and Heintz, F. and Hinton, G. and Jennings, N. and Leavy, S. and Ludermir, T. and Marda, V. and Margetts, H. and McDermid, J. and Munga, J. and Narayanan, A. and Nelson, A. and Neppel, C. and Ramchurn, S. D. and Russell, S. and Schaake, M. and Schölkopf, B. and Soto, A. and Tiedrich, L. and Varoquaux, G. and Yao, A. and Zhang, Y.-Q. and Aguirre, L. A. and Ajala, O. and Albalawi, F. and AlMalek, N. and Busch, C. and Collas, J. and de Carvalho, A. C. P. de L. F. and Gill, A. and Hatip, A. H. and Heikkilä, J. and Johnson, C. and Jolly, G. and Katzir, Z. and Kerema, M. N. and Kitano, H. and Krüger, A. and Lee, K. M. and López Portillo, J. R. and McLysaght, A. and Molchanovskyi, O. and Monti, A. and Nemer, M. and Oliver, N. and Pezoa, R. and Plonk, A. and Ravindran, B. and Riza, H. and Rugege, C. and Sheikh, H. and Wong, D. and Zeng, Y. and Zhu, L. and Privitera, D. and Mindermann, S.},
	year = {2026},
}

@misc{manning_how_2026,
	type = {Working {Paper}},
	series = {Working {Paper} {Series}},
	title = {How {Adaptable} {Are} {American} {Workers} to {AI}-{Induced} {Job} {Displacement}?},
	url = {https://www.nber.org/papers/w34705},
	doi = {10.3386/w34705},
	urldate = {2026-02-11},
	publisher = {National Bureau of Economic Research},
	author = {Manning, Sam J. and Aguirre, Tomás},
	month = jan,
	year = {2026},
	doi = {10.3386/w34705},
}

@misc{fronsdal_petri_2026,
	title = {Petri 2.0: {New} scenarios, new model comparisons, and improved eval-awareness mitigations},
	url = {https://alignment.anthropic.com/2026/petri-v2/},
	author = {Fronsdal, Kai and Michala, Jonathan and Bowman, Sam},
	year = {2026},
}

@misc{starace_paperbench_2025,
	title = {{PaperBench}: {Evaluating} {AI}'s {Ability} to {Replicate} {AI} {Research}},
	shorttitle = {{PaperBench}},
	url = {http://arxiv.org/abs/2504.01848},
	doi = {10.48550/arXiv.2504.01848},
	urldate = {2026-01-28},
	publisher = {arXiv},
	author = {Starace, Giulio and Jaffe, Oliver and Sherburn, Dane and Aung, James and Chan, Jun Shern and Maksin, Leon and Dias, Rachel and Mays, Evan and Kinsella, Benjamin and Thompson, Wyatt and Heidecke, Johannes and Glaese, Amelia and Patwardhan, Tejal},
	month = apr,
	year = {2025},
	note = {arXiv:2504.01848 [cs]},
}

@misc{nielsen_notes_2024,
	title = {Notes on {Differential} {Technological} {Development}},
	url = {https://michaelnotebook.com/dtd/},
	urldate = {2026-01-28},
	author = {Nielsen, Michael},
	year = {2024},
}

@misc{dellacqua_navigating_2023,
	address = {Rochester, NY},
	type = {{SSRN} {Scholarly} {Paper}},
	title = {Navigating the {Jagged} {Technological} {Frontier}: {Field} {Experimental} {Evidence} of the {Effects} of {AI} on {Knowledge} {Worker} {Productivity} and {Quality}},
	shorttitle = {Navigating the {Jagged} {Technological} {Frontier}},
	url = {https://papers.ssrn.com/abstract=4573321},
	doi = {10.2139/ssrn.4573321},
	language = {en},
	urldate = {2026-01-28},
	publisher = {Social Science Research Network},
	author = {Dell'Acqua, Fabrizio and McFowland III, Edward and Mollick, Ethan R. and Lifshitz-Assaf, Hila and Kellogg, Katherine and Rajendran, Saran and Krayer, Lisa and Candelon, François and Lakhani, Karim R.},
	month = sep,
	year = {2023},
}

@misc{macdiarmid_natural_2025,
	title = {Natural {Emergent} {Misalignment} from {Reward} {Hacking} in {Production} {RL}},
	url = {http://arxiv.org/abs/2511.18397},
	doi = {10.48550/arXiv.2511.18397},
	urldate = {2026-01-28},
	publisher = {arXiv},
	author = {MacDiarmid, Monte and Wright, Benjamin and Uesato, Jonathan and Benton, Joe and Kutasov, Jon and Price, Sara and Bouscal, Naia and Bowman, Sam and Bricken, Trenton and Cloud, Alex and Denison, Carson and Gasteiger, Johannes and Greenblatt, Ryan and Leike, Jan and Lindsey, Jack and Mikulik, Vlad and Perez, Ethan and Rodrigues, Alex and Thomas, Drake and Webson, Albert and Ziegler, Daniel and Hubinger, Evan},
	month = nov,
	year = {2025},
	note = {arXiv:2511.18397 [cs]},
}

@techreport{buterin_my_2023,
	title = {My techno-optimism},
	author = {Buterin, Vitalik},
	year = {2023},
}

@misc{erdil_most_2025,
	title = {Most {AI} value will come from broad automation, not from {R}\&{D}},
	url = {https://epoch.ai/gradient-updates/most-ai-value-will-come-from-broad-automation-not-from-r-d},
	author = {Erdil, Ege and Barnett, Matthew},
	year = {2025},
}

@misc{fronsdal_petri_2025,
	title = {Petri: {Parallel} exploration of risky interactions},
	url = {https://github.com/safety-research/petri},
	author = {Fronsdal, Kai and Gupta, Isha and Sheshadri, Abhay and Michala, Jonathan and McAleer, Stephen and Wang, Rowan and Price, Sara and Bowman, Sam},
	year = {2025},
}

@misc{gans_oring_2026,
	type = {Working {Paper}},
	series = {Working {Paper} {Series}},
	title = {O-{Ring} {Automation}},
	url = {https://www.nber.org/papers/w34639},
	doi = {10.3386/w34639},
	urldate = {2026-01-28},
	publisher = {National Bureau of Economic Research},
	author = {Gans, Joshua S. and Goldfarb, Avi},
	month = jan,
	year = {2026},
	doi = {10.3386/w34639},
}

@misc{you_most_2025,
	title = {Most of {OpenAI}’s 2024 compute went to experiments},
	url = {https://epoch.ai/data-insights/openai-compute-spend},
	author = {You, Josh},
	year = {2025},
}

@misc{becker_measuring_2025,
	title = {Measuring the {Impact} of {Early}-2025 {AI} on {Experienced} {Open}-{Source} {Developer} {Productivity}},
	url = {http://arxiv.org/abs/2507.09089},
	doi = {10.48550/arXiv.2507.09089},
	urldate = {2026-01-28},
	publisher = {arXiv},
	author = {Becker, Joel and Rush, Nate and Barnes, Elizabeth and Rein, David},
	month = jul,
	year = {2025},
	note = {arXiv:2507.09089 [cs]},
}

@misc{epochai_introducing_2024,
	title = {Introducing {Epoch} {AI}’s {AI} benchmarking hub},
	url = {https://epoch.ai/blog/introducing-benchmarks-dashboard},
	language = {en},
	urldate = {2026-01-28},
	journal = {Epoch AI},
	author = {{Epoch AI}},
	month = nov,
	year = {2024},
}

@misc{owen_interviewing_2024,
	title = {Interviewing {AI} researchers on automation of {AI} {R}\&{D}},
	url = {https://epoch.ai/blog/interviewing-ai-researchers-on-automation-of-ai-rnd},
	author = {Owen, David},
	year = {2024},
}

@misc{cotroneo_humanwritten_2025,
	title = {Human-{Written} vs. {AI}-{Generated} {Code}: {A} {Large}-{Scale} {Study} of {Defects}, {Vulnerabilities}, and {Complexity}},
	shorttitle = {Human-{Written} vs. {AI}-{Generated} {Code}},
	url = {http://arxiv.org/abs/2508.21634},
	doi = {10.48550/arXiv.2508.21634},
	urldate = {2026-01-28},
	publisher = {arXiv},
	author = {Cotroneo, Domenico and Improta, Cristina and Liguori, Pietro},
	month = aug,
	year = {2025},
	note = {arXiv:2508.21634 [cs]},
}

@misc{chan_mlebench_2025,
	title = {{MLE}-bench: {Evaluating} {Machine} {Learning} {Agents} on {Machine} {Learning} {Engineering}},
	shorttitle = {{MLE}-bench},
	url = {http://arxiv.org/abs/2410.07095},
	doi = {10.48550/arXiv.2410.07095},
	urldate = {2026-01-28},
	publisher = {arXiv},
	author = {Chan, Jun Shern and Chowdhury, Neil and Jaffe, Oliver and Aung, James and Sherburn, Dane and Mays, Evan and Starace, Giulio and Liu, Kevin and Maksin, Leon and Patwardhan, Tejal and Weng, Lilian and Mądry, Aleksander},
	month = feb,
	year = {2025},
	note = {arXiv:2410.07095 [cs]},
}

@misc{metr_measuring_2025,
	title = {Measuring {AI} ability to complete long tasks},
	url = {https://metr.org/blog/2025-03-19-measuring-ai-ability-to-complete-long-tasks/},
	author = {{METR}},
	month = mar,
	year = {2025},
}

@misc{marin_marin_2025,
	title = {Marin {32B} {Retrospective}},
	url = {https://marin.readthedocs.io/en/latest/reports/marin-32b-retro/},
	urldate = {2026-01-28},
	author = {{Marin}},
	year = {2025},
}

@techreport{openai_gpt5_2025,
	title = {{GPT}-5 {System} {Card}},
	url = {https://cdn.openai.com/gpt-5-system-card.pdf},
	author = {{OpenAI}},
	month = aug,
	year = {2025},
}

@misc{brynjolfsson_generative_2023,
	type = {Working {Paper}},
	series = {Working {Paper} {Series}},
	title = {Generative {AI} at {Work}},
	url = {https://www.nber.org/papers/w31161},
	doi = {10.3386/w31161},
	urldate = {2026-01-28},
	publisher = {National Bureau of Economic Research},
	author = {Brynjolfsson, Erik and Li, Danielle and Raymond, Lindsey R.},
	month = apr,
	year = {2023},
	doi = {10.3386/w31161},
}

@techreport{googledeepmind_gemini_2025,
	title = {Gemini 3 {Pro} {Model} {Card}},
	url = {https://storage.googleapis.com/deepmind-media/Model-Cards/Gemini-3-Pro-Model-Card.pdf},
	author = {{Google DeepMind}},
	month = nov,
	year = {2025},
}

@misc{erdil_gate_2025,
	title = {{GATE}: {An} integrated assessment model for {AI} automation},
	url = {https://arxiv.org/abs/2503.04941},
	author = {Erdil, Ege and Potlogea, Andrei and Besiroglu, Tamay and Roldan, Edu and Ho, Anson and Sevilla, Jaime and Barnett, Matthew and Vrzla, Matej and Sandler, Robert},
	year = {2025},
	note = {arXiv: 2503.04941 [econ.GN]},
}

@misc{meinke_frontier_2025,
	title = {Frontier {Models} are {Capable} of {In}-context {Scheming}},
	url = {http://arxiv.org/abs/2412.04984},
	doi = {10.48550/arXiv.2412.04984},
	urldate = {2026-01-28},
	publisher = {arXiv},
	author = {Meinke, Alexander and Schoen, Bronson and Scheurer, Jérémy and Balesni, Mikita and Shah, Rusheb and Hobbhahn, Marius},
	month = jan,
	year = {2025},
	note = {arXiv:2412.04984 [cs]},
}

@misc{karger_forecastbench_2025,
	title = {{ForecastBench}: {A} {Dynamic} {Benchmark} of {AI} {Forecasting} {Capabilities}},
	shorttitle = {{ForecastBench}},
	url = {http://arxiv.org/abs/2409.19839},
	doi = {10.48550/arXiv.2409.19839},
	urldate = {2026-01-28},
	publisher = {arXiv},
	author = {Karger, Ezra and Bastani, Houtan and Yueh-Han, Chen and Jacobs, Zachary and Halawi, Danny and Zhang, Fred and Tetlock, Philip E.},
	month = feb,
	year = {2025},
	note = {arXiv:2409.19839 [cs]},
}

@article{noy_experimental_2023,
	title = {Experimental evidence on the productivity effects of generative artificial intelligence},
	volume = {381},
	url = {https://www.science.org/doi/10.1126/science.adh2586},
	doi = {10.1126/science.adh2586},
	number = {6654},
	urldate = {2026-01-28},
	journal = {Science},
	publisher = {American Association for the Advancement of Science},
	author = {Noy, Shakked and Zhang, Whitney},
	month = jul,
	year = {2023},
	pages = {187--192},
}

@misc{zhu_establishing_2025,
	title = {Establishing {Best} {Practices} for {Building} {Rigorous} {Agentic} {Benchmarks}},
	url = {http://arxiv.org/abs/2507.02825},
	doi = {10.48550/arXiv.2507.02825},
	urldate = {2026-01-28},
	publisher = {arXiv},
	author = {Zhu, Yuxuan and Jin, Tengjun and Pruksachatkun, Yada and Zhang, Andy and Liu, Shu and Cui, Sasha and Kapoor, Sayash and Longpre, Shayne and Meng, Kevin and Weiss, Rebecca and Barez, Fazl and Gupta, Rahul and Dhamala, Jwala and Merizian, Jacob and Giulianelli, Mario and Coppock, Harry and Ududec, Cozmin and Sekhon, Jasjeet and Steinhardt, Jacob and Kellerman, Antony and Schwettmann, Sarah and Zaharia, Matei and Stoica, Ion and Liang, Percy and Kang, Daniel},
	month = jul,
	year = {2025},
	note = {arXiv:2507.02825 [cs]
version: 2},
}

@article{betley_emergent_2026,
	title = {Emergent {Misalignment}: {Narrow} finetuning can produce broadly misaligned {LLMs}},
	volume = {649},
	issn = {0028-0836, 1476-4687},
	shorttitle = {Emergent {Misalignment}},
	url = {http://arxiv.org/abs/2502.17424},
	doi = {10.1038/s41586-025-09937-5},
	number = {8097},
	urldate = {2026-01-28},
	journal = {Nature},
	author = {Betley, Jan and Tan, Daniel and Warncke, Niels and Sztyber-Betley, Anna and Bao, Xuchan and Soto, Martín and Labenz, Nathan and Evans, Owain},
	month = jan,
	year = {2026},
	note = {arXiv:2502.17424 [cs]},
	pages = {584--589},
}

@misc{sandbrink_differential_2022,
	address = {Rochester, NY},
	type = {{SSRN} {Scholarly} {Paper}},
	title = {Differential technology development: {An} innovation governance consideration for navigating technology risks},
	shorttitle = {Differential technology development},
	url = {https://papers.ssrn.com/abstract=4213670},
	doi = {10.2139/ssrn.4213670},
	language = {en},
	urldate = {2026-01-28},
	publisher = {Social Science Research Network},
	author = {Sandbrink, Jonas and Hobbs, Hamish and Swett, Jacob and Dafoe, Allan and Sandberg, Anders},
	month = sep,
	year = {2022},
}

@techreport{googledeepmind_frontier_2025,
	title = {Frontier {Safety} {Framework}},
	number = {Version 3.0},
	author = {{Google DeepMind}},
	year = {2025},
}

@article{narayanan_ai_2025,
	title = {{AI} as normal technology},
	volume = {25},
	journal = {Knight First Amendment Institute},
	author = {Narayanan, Arvind and Kapoor, Sayash},
	year = {2025},
}

@misc{kamilelukosiute_design_2025,
	title = {Design for the defenders you care about or risk being useless},
	url = {https://kamilelukosiute.com/llms/Design+for+the+defenders+you+care+about+or+risk+being+useless},
	language = {en},
	urldate = {2026-01-28},
	author = {{Kamilė Lukošiūtė}},
	year = {2025},
}

@article{perrigo_demis_2025,
	title = {Demis {Hassabis} {Is} {Preparing} for {AI}’s {Endgame}},
	url = {https://time.com/7277608/demis-hassabis-interview-time100-2025/},
	language = {en},
	urldate = {2025-12-17},
	journal = {TIME},
	author = {Perrigo, Billy},
	month = apr,
	year = {2025},
}

@misc{ward_ctrlaltdeceit_2025,
	title = {{CTRL}-{ALT}-{DECEIT}: {Sabotage} {Evaluations} for {Automated} {AI} {R}\&{D}},
	shorttitle = {{CTRL}-{ALT}-{DECEIT}},
	url = {http://arxiv.org/abs/2511.09904},
	doi = {10.48550/arXiv.2511.09904},
	urldate = {2026-01-28},
	publisher = {arXiv},
	author = {Ward, Francis Rhys and Weij, Teun van der and Gábor, Hanna and Martin, Sam and Moreno, Raja Mehta and Lidar, Harel and Makower, Louis and Jodrell, Thomas and Robson, Lauren},
	month = nov,
	year = {2025},
	note = {arXiv:2511.09904 [cs]},
}

@misc{bai_constitutional_2022,
	title = {Constitutional {AI}: {Harmlessness} from {AI} {Feedback}},
	shorttitle = {Constitutional {AI}},
	url = {http://arxiv.org/abs/2212.08073},
	doi = {10.48550/arXiv.2212.08073},
	urldate = {2026-01-28},
	publisher = {arXiv},
	author = {Bai, Yuntao and Kadavath, Saurav and Kundu, Sandipan and Askell, Amanda and Kernion, Jackson and Jones, Andy and Chen, Anna and Goldie, Anna and Mirhoseini, Azalia and McKinnon, Cameron and Chen, Carol and Olsson, Catherine and Olah, Christopher and Hernandez, Danny and Drain, Dawn and Ganguli, Deep and Li, Dustin and Tran-Johnson, Eli and Perez, Ethan and Kerr, Jamie and Mueller, Jared and Ladish, Jeffrey and Landau, Joshua and Ndousse, Kamal and Lukosuite, Kamile and Lovitt, Liane and Sellitto, Michael and Elhage, Nelson and Schiefer, Nicholas and Mercado, Noemi and DasSarma, Nova and Lasenby, Robert and Larson, Robin and Ringer, Sam and Johnston, Scott and Kravec, Shauna and Showk, Sheer El and Fort, Stanislav and Lanham, Tamera and Telleen-Lawton, Timothy and Conerly, Tom and Henighan, Tom and Hume, Tristan and Bowman, Samuel R. and Hatfield-Dodds, Zac and Mann, Ben and Amodei, Dario and Joseph, Nicholas and McCandlish, Sam and Brown, Tom and Kaplan, Jared},
	month = dec,
	year = {2022},
	note = {arXiv:2212.08073 [cs]},
}

@misc{tamkin_clio_2024,
	title = {Clio: {Privacy}-{Preserving} {Insights} into {Real}-{World} {AI} {Use}},
	shorttitle = {Clio},
	url = {http://arxiv.org/abs/2412.13678},
	doi = {10.48550/arXiv.2412.13678},
	urldate = {2026-01-28},
	publisher = {arXiv},
	author = {Tamkin, Alex and McCain, Miles and Handa, Kunal and Durmus, Esin and Lovitt, Liane and Rathi, Ankur and Huang, Saffron and Mountfield, Alfred and Hong, Jerry and Ritchie, Stuart and Stern, Michael and Clarke, Brian and Goldberg, Landon and Sumers, Theodore R. and Mueller, Jared and McEachen, William and Mitchell, Wes and Carter, Shan and Clark, Jack and Kaplan, Jared and Ganguli, Deep},
	month = dec,
	year = {2024},
	note = {arXiv:2412.13678 [cs]},
}

@misc{clymer_bare_2025,
	title = {Bare {Minimum} {Mitigations} for {Autonomous} {AI} {Development}},
	url = {http://arxiv.org/abs/2504.15416},
	doi = {10.48550/arXiv.2504.15416},
	urldate = {2026-01-28},
	publisher = {arXiv},
	author = {Clymer, Joshua and Duan, Isabella and Cundy, Chris and Duan, Yawen and Heide, Fynn and Lu, Chaochao and Mindermann, Sören and McGurk, Conor and Pan, Xudong and Siddiqui, Saad and Wang, Jingren and Yang, Min and Zhan, Xianyuan},
	month = apr,
	year = {2025},
	note = {arXiv:2504.15416 [cs]},
}

@misc{inglis_controlarena_2025,
	title = {{ControlArena}},
	url = {https://github.com/UKGovernmentBEIS/control-arena},
	author = {Inglis, Rogan and Matthews, Ollie and Tracy, Tyler and Makins, Oliver and Catling, Tom and Cooper Stickland, Asa and Faber-Espensen, Rasmus and O'Connell, Daniel and Heller, Myles and Brandao, Miguel and Hanson, Adam and Mani, Arathi and Korbak, Tomek and Michelfeit, Jan and Bansal, Dishank and Bark, Tomas and Canal, Chris and Griffin, Charlie and Wang, Jasmine and Cooney, Alan},
	year = {2025},
}

@inproceedings{sevilla_compute_2022,
	title = {Compute trends across three eras of machine learning},
	doi = {10.1109/IJCNN55064.2022.9891914},
	booktitle = {2022 international joint conference on neural networks ({IJCNN})},
	author = {Sevilla, Jaime and Heim, Lennart and Ho, Anson and Besiroglu, Tamay and Hobbhahn, Marius and Villalobos, Pablo},
	year = {2022},
	pages = {1--8},
}

@misc{greenblatt_alignment_2024,
	title = {Alignment faking in large language models},
	url = {http://arxiv.org/abs/2412.14093},
	doi = {10.48550/arXiv.2412.14093},
	urldate = {2026-01-28},
	publisher = {arXiv},
	author = {Greenblatt, Ryan and Denison, Carson and Wright, Benjamin and Roger, Fabien and MacDiarmid, Monte and Marks, Sam and Treutlein, Johannes and Belonax, Tim and Chen, Jack and Duvenaud, David and Khan, Akbir and Michael, Julian and Mindermann, Sören and Perez, Ethan and Petrini, Linda and Uesato, Jonathan and Kaplan, Jared and Shlegeris, Buck and Bowman, Samuel R. and Hubinger, Evan},
	month = dec,
	year = {2024},
	note = {arXiv:2412.14093 [cs]},
}

@misc{barnett_algorithmic_2025,
	title = {Algorithmic progress likely spurs more spending on compute, not less},
	url = {https://epoch.ai/gradient-updates/algorithmic-progress-likely-spurs-more-spending-on-compute-not-less},
	author = {Barnett, Matthew},
	year = {2025},
}

@misc{ho_algorithmic_2024,
	title = {Algorithmic progress in language models},
	author = {Ho, Anson and Besiroglu, Tamay and Erdil, Ege and Owen, David and Rahman, Robi and Guo, Zifan Carl and Atkinson, David and Thompson, Neil and Sevilla, Jaime},
	year = {2024},
	note = {arXiv: 2403.05812 [cs.CL]},
}

@misc{schoenegger_aiaugmented_2024,
	title = {{AI}-{Augmented} {Predictions}: {LLM} {Assistants} {Improve} {Human} {Forecasting} {Accuracy}},
	shorttitle = {{AI}-{Augmented} {Predictions}},
	url = {http://arxiv.org/abs/2402.07862},
	doi = {10.48550/arXiv.2402.07862},
	urldate = {2026-01-28},
	publisher = {arXiv},
	author = {Schoenegger, Philipp and Park, Peter S. and Karger, Ezra and Trott, Sean and Tetlock, Philip E.},
	month = aug,
	year = {2024},
	note = {arXiv:2402.07862 [cs]},
}

@misc{aifuturesproject_ai_2026,
	title = {{AI} {Futures} {Model}},
	url = {https://www.aifuturesmodel.com/},
	language = {en},
	urldate = {2026-01-28},
	author = {{AI Futures Project}},
	year = {2026},
}

@article{davidson_aienabled_2025,
	title = {{AI}-enabled coups: {How} a small group could use {AI} to seize power},
	url = {https://www.forethought.org/research/ai-enabled-coups-how-a-small-group-could-use-ai-to-seize-power},
	author = {Davidson, Tom and Finnveden, Lukas and Hadshar, Rose},
	year = {2025},
}

@article{vaintrob_ai_2025,
	title = {{AI} tools for existential security},
	url = {https://www.forethought.org/research/ai-tools-for-existential-security},
	author = {Vaintrob, Lizka and Cotton-Barratt, Owen},
	year = {2025},
}

@misc{kembery_ai_2025,
	title = {{AI} {Resilience}},
	url = {https://airesilience.net/},
	language = {en},
	urldate = {2026-01-28},
	author = {Kembery, Eddie and Ammann, Nora},
	year = {2025},
}

@misc{metr_ai_2025,
	title = {{AI} models can be dangerous before public deployment},
	url = {https://metr.org/blog/2025-01-17-ai-models-dangerous-before-public-deployment/},
	author = {{METR}},
	month = jan,
	year = {2025},
}

@misc{stix_ai_2025,
	title = {{AI} {Behind} {Closed} {Doors}: a {Primer} on {The} {Governance} of {Internal} {Deployment}},
	shorttitle = {{AI} {Behind} {Closed} {Doors}},
	url = {http://arxiv.org/abs/2504.12170},
	doi = {10.48550/arXiv.2504.12170},
	urldate = {2026-01-28},
	publisher = {arXiv},
	author = {Stix, Charlotte and Pistillo, Matteo and Sastry, Girish and Hobbhahn, Marius and Ortega, Alejandro and Balesni, Mikita and Hallensleben, Annika and Goldowsky-Dill, Nix and Sharkey, Lee},
	month = apr,
	year = {2025},
	note = {arXiv:2504.12170 [cs]},
}

@misc{korbak_sketch_2025,
	title = {A sketch of an {AI} control safety case},
	url = {http://arxiv.org/abs/2501.17315},
	doi = {10.48550/arXiv.2501.17315},
	urldate = {2026-01-28},
	publisher = {arXiv},
	author = {Korbak, Tomek and Clymer, Joshua and Hilton, Benjamin and Shlegeris, Buck and Irving, Geoffrey},
	month = jan,
	year = {2025},
	note = {arXiv:2501.17315 [cs]},
}

@misc{stackoverflow_2025_2025,
	title = {2025 {Developer} {Survey}},
	url = {https://survey.stackoverflow.co/2025/ai},
	language = {en},
	urldate = {2026-01-28},
	author = {{Stack Overflow}},
	year = {2025},
}
\bibliographystyle{tmlr}

% \appendix
% \section{Appendix}
% You may include other additional sections here.

\end{document}